\documentclass[%
 reprint,
 amsmath,amssymb,
 aps,
]{revtex4-2}

\usepackage{graphicx}
\usepackage{dcolumn}
\usepackage{bm}
\usepackage{hyperref}
\usepackage{xcolor}
\usepackage[percent]{overpic}

\newcommand{\R}{\mathbb{R}}

\newcommand{\og}{\Omega_{G}}
\newcommand{\oo}{\Omega_{0}}

\newcommand{\mlow}{M^{\text{low}}}

\begin{document}

\preprint{APS/123-QED}

\title{Shepherding control and herdability in complex multiagent systems}

\author{Andrea Lama}
\email{andrea.lama-ssm@unina.it}
\affiliation{Scuola Superiore Meridionale, Naples, Italy}

\author{Mario di Bernardo}%
 \email{mario.dibernardo@unina.it}
\affiliation{Department of Electrical Engineering and ICT, University of Naples Federico II, Italy
}
\altaffiliation[Also at ]{Scuola Superiore Meridionale, Naples, Italy.}

\date{\today}

\begin{abstract}
We study the shepherding control problem where a group of ``herders’' need to orchestrate their collective behaviour in order to steer the dynamics of a group of ``target’' agents towards a desired goal. We relax the strong assumptions of targets showing cohesive collective behavior in the absence of the herders, and herders owning global sensing capabilities. We find  scaling laws linking the number of targets and minimum herders needed, and we unveil the existence of a critical threshold of the density of the targets, below which the number of herders needed for success significantly increases. We explain the existence of such a threshold in terms of the percolation of a suitably defined herdability graph and support our numerical evidence by deriving and analysing a PDE describing the herders dynamics in a simplified one-dimensional setting. Extensive numerical experiments validate our methodology.
\end{abstract}

\maketitle

In various physical contexts, a group of agents (the {\em herders}) must cooperate and self-organize in order to orchestrate the emergence of some desired collective behavior in another group (the {\em targets}), which would naturally act differently. Examples include shepherd dogs herding sheep towards a location \cite{strombom2014solving}, predators like dolphins corralling prey \cite{haque2011biologically,bailey2013group}, and robotic systems managing environmental pollutants \cite{zahugi2013oil} or guiding agents to safety \cite{sebastian2022adaptive,pierson2017controlling,licitra2019single,long2020comprehensive,yuan2023multi}. This is known as the ``shepherding'' control problem in the control theoretic and robotics literature, where typically artificial herders need to drive targets to a specified area \cite{long2020comprehensive}. Previous research has primarily been focused on one herder with multiple targets \cite{licitra2019single,ko2020asymptotic, ranganathan2022optimal}, with less emphasis on the case of multiple herders \cite{sebastian2022adaptive,auletta2022herding,pierson2017controlling}. 

Notably, when herders are outnumbered, solutions often assume targets exhibit cohesive behavior, like sheep flocking together \cite{ko2020asymptotic,strombom2014solving,fujioka2017comparison,sebastian2022adaptive,van2023steering}, allowing herders to leverage this for efficient problem-solving \cite{ranganathan2022optimal,sebastian2022adaptive,ko2020asymptotic}.
Relaxing this assumption makes the problem much more cumbersome to solve theoretically, as recently noted in \cite{ko2020asymptotic}, and is also unrealistic in many applications such as environmental cleanup via multi-robot systems \cite{zahugi2013oil} or the confinement of microbial populations \cite{massana2022rectification} where target agents (pollutant particles or bacteria) do not necessarily fulfill this hypothesis.

There is also a second strong and, most importantly, unrealistic assumption that most of the current solutions often adopt: that the herders possess unlimited sensing capabilities, i.e. that they all know the positions of all other herders and all targets in the region of interest \cite{auletta2022herding}.  
 
Moreover, as noted  independently in the recent literature, e.g. \cite{ranganathan2022optimal,auletta2022herding}, most existing approaches adopt centralized or distributed strategies that do not exploit a crucial feature of complex systems. Specifically, effective shepherding control should not be pre-engineered but should emerge from herders following simple local rules in their interactions with targets and each other, leading to collective behavior suited for the shepherding task. A striking example is that described in \cite{nalepka2019pnas}, where a phenomenological model is used to describe the emergent collective behaviour that two or more humans show when asked to solve the shepherding problem in a virtual reality setting (e.g. starting oscillating around the targets to contain them).

In this Letter, contrary to the existing literature, we simultaneously remove both of the assumptions mentioned earlier. Our investigation aims to determine if, and under what ``herdability'' conditions, multiple cooperating herders -- operating solely based on local information -- can effectively shepherd a group of target agents towards a desired state. To this aim, we propose and analyse a minimal model, general enough to comprise all the crucial features of the problem in the presence of limited sensing capabilities of the herders and the absence of any inherent collective behavior among the targets.



Our approach is fundamentally different from typical studies  on active-passive particle systems, see e.g. \cite{stenhammar2015activity, dolai2018phase,forgacs2021active}, where the emergence is also observed of spontaneous behaviour similar to shepherding. It is also distinct from previous research on bipartite systems for understanding prey-predator and foraging behaviors \cite{mohapatra2019confined}. In particular, in the problem we study herders actively make decisions on what targets to select and maneuver incorporating a feedback mechanism based on their proximity to themselves and the goal region. This unique integration of feedback control theory into physics-inspired models marks our study as distinct in the field of complex systems control. Our goal is to engineer the collective behaviour of a complex multiagent system (the herders)  in order for another group of agents (the targets) to perform a desired task and solve a distributed control problem; an aspect that has been rarely considered in the vast literature on control and controllability of complex systems (see e.g. \cite{d2023controlling} and references therein). 

\begin{figure}[tb]
\begin{overpic}[scale=0.85]{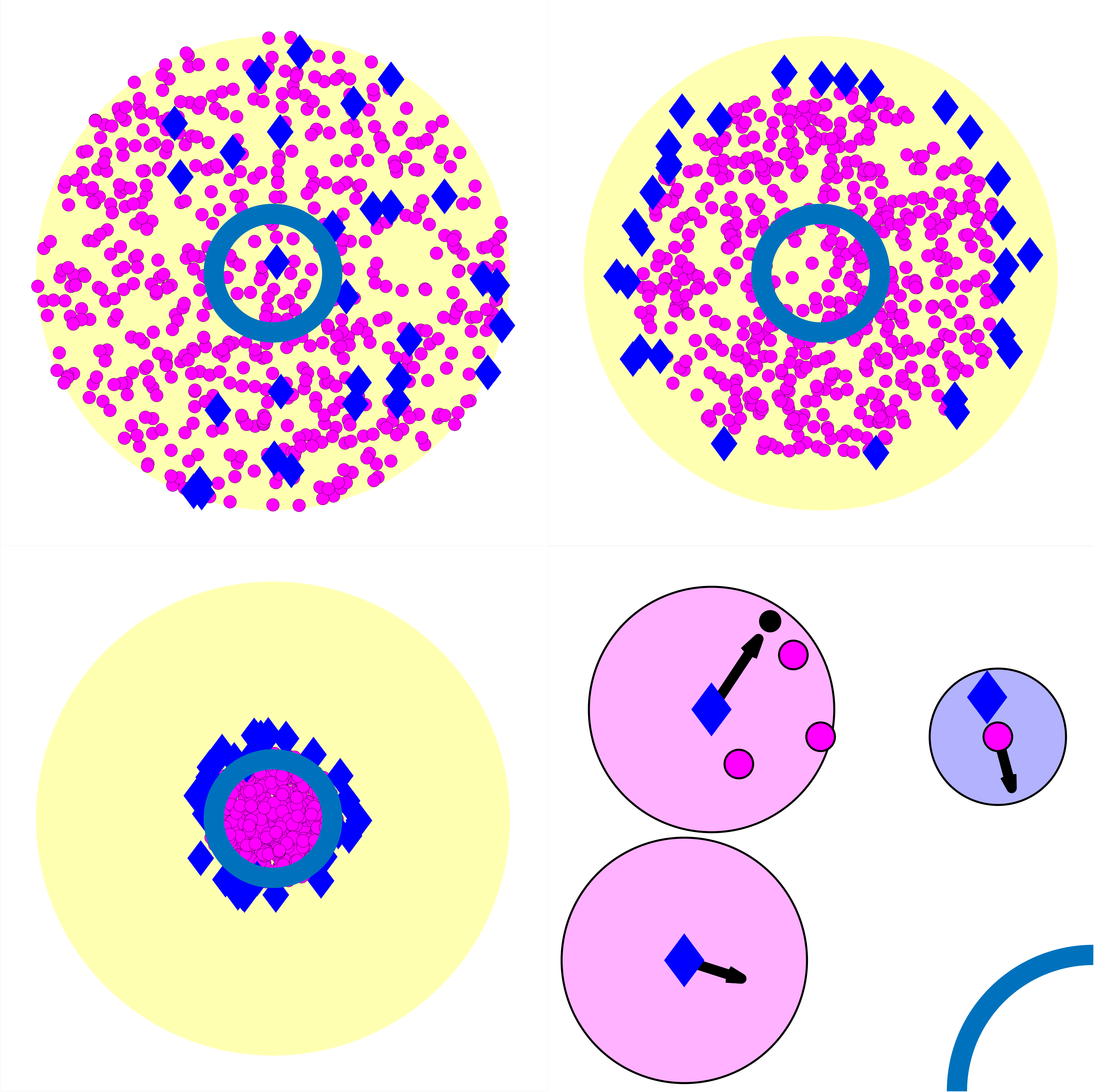}
 \put(0,100){\makebox(0,0){\textbf{(a)}}}
 \put(55,100){\makebox(0,0){\textbf{(b)}}}
 \put(0,50){\makebox(0,0){\textbf{(c)}}}
 \put(55,50){\makebox(0,0){\textbf{(d)}}}
 \put(58,15){\makebox(0,0){\textbf{$H_1$}}}
 \put(63,5){\makebox(0,0){\textbf{$\xi$}}}
 \put(60,39){\makebox(0,0){\textbf{$H_2$}}}
  \put(90,42){\makebox(0,0){\textbf{$H_3$}}}
  \put(95,32){\makebox(0,0){\textbf{$\lambda$}}}
    \put(95,5){\makebox(0,0){\textbf{$\og$}}}
\end{overpic}

\caption{\label{fig:PRL_HCP}{Representative snapshots of the system configuration (with herders represented by blue diamonds and targets by magenta circles) at (a) the initial time $t=0$ with the agents uniformly distributed in $\oo$ (yellow shaded disk), (b) at an intermediate time during shepherding control when herders surround all targets  and (c) when the task is successfully achieved with all the targets in $\og$ (dark blue circle). (d) Schematic of the herders' and targets' sensing (magenta shaded disks) and repulsion (blue shaded disk) regions of radius $\xi$ and $\lambda$ respectively. The solid black arrows represent the direction of motion of the herders when moving in the absence of nearby targets ($\textbf{H}_1$), selecting the target to chase $\textbf{T}^*$ with the largest distance from to goal ($\textbf{H}_2$), or when the herder pushes a selected target towards the goal region ($\textbf{H}_3$).}}
\end{figure}
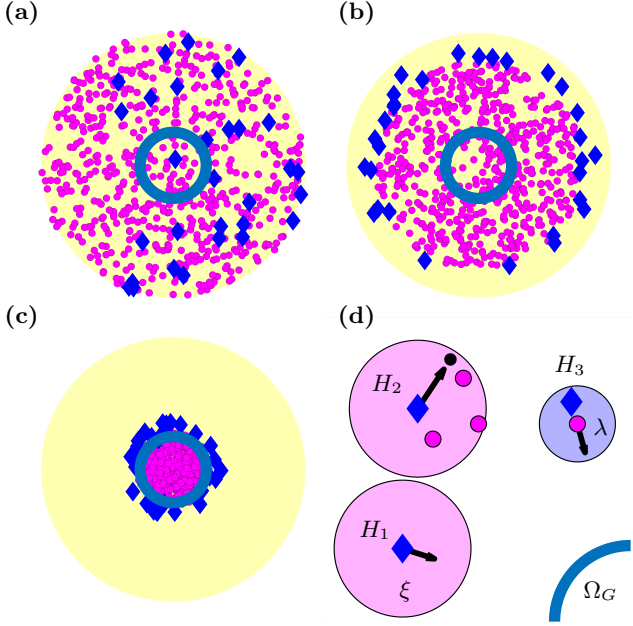

We consider the shepherding problem in $\mathbb{R}^2$ (see Fig. \hyperref[fig:HCP]{\ref{fig:PRL_HCP}a}), where $N$ herders have to corral $M$ targets to a goal region $\og$. We assume that both the herders and the targets are initially randomly and uniformly distributed in a circle $\Omega_0$ of radius $R$, and that the $\og$ region is a circle of radius $r^*<R$; both $\Omega_0$ and $\og$ being centered around the origin.
Let $\mathbf{H}\in \R ^{2N}$ be the vector of the herders' positions $\mathbf{H}=\left[\mathbf{H}_1,\,\mathbf{H}_2,\,...,\,\mathbf{H}_N\right]$ with $ \mathbf{H}_i \in \R^2$ being the Cartesian coordinates of the $i$-th herder, $i=1,...,N$, and $\mathbf{T}\in \R ^{2M}$ the vector of the targets' positions  $\mathbf{T}=\left[\mathbf{T}_1,\,\mathbf{T}_2,\,...,\,\mathbf{T}_M\right]$, with $\mathbf{T}_a \in \R^2$  being the Cartesian coordinates of target $a$, $a=1,...,M$.

We assume the targets do not exhibt any type of cohesive collective behaviour with their dynamics being described by the following overdamped Langevin equation:
\begin{equation}
    \dot{\textbf{T}}_a =\sqrt{2D}\bm{\mathcal{N}}+\beta\sum_{i \in \mathcal{N}_a}\left(\lambda-\left \vert \textbf{d}_{ia} \right \vert\right )\hat{\textbf{d}}_{ia}
    \label{eqn:targets}
\end{equation}
where, analogously to what typically considered in the literature on soft matter, e.g. \cite{dolai2018phase,casiulis2021self,forgacs2021active}, $\bm{\mathcal{N}}$ is white Gaussian noise, $\beta$ and $D$ are positive constants, $\textbf{d}_{ia}=\textbf{H}_i-\textbf{T}_a$ is the distance between herder $i$ and target $a$, $\lambda>0$ is the radius of the region where targets are repelled by nearby herders, and $\mathcal{N}_a$ represents the set of indexes of all the herders, if any, whose positions are such that $|\textbf{d}_{ia}|\leq \lambda$. Note that  $\beta\lambda\equiv v_T$ is the maximum escaping speed of a target due to the presence of a nearby herder and that we assume $\beta\lambda^2 \gg D$ so that the harmonic repulsive action eventually exerted by the herders onto the targets dominates over their own Brownian dynamics 

We model the dynamics of the herders as made of two mutually exclusive terms, one capturing their own dynamics and the other their interaction with the targets, see e.g. \cite{romanczuk2009collective}. Specifically, we set
\begin{equation}
    \dot{\textbf{H}}_i=(1-\eta_i)\textbf{F}_i(\textbf{H}_i,\textbf{r}^*) + \eta_i \textbf{I}_i(\textbf{T},\textbf{H},\xi)
    \label{eqn:herders}
\end{equation}
where  $\eta_i=\eta_i(\textbf{T},\textbf{H},\xi)$ is an indicator function activating when herder $i$ has at least one target to chase in its sensing region of radius $\xi$, 
$\textbf{F}_i(\textbf{H}_i,r^*)$ describes the herder's own dynamics when it is not chasing any targets while $\textbf{I}_i(\textbf{T},\textbf{H},\xi)$ is a feedback term capturing the herder's reaction to the presence of targets in its sensing region.

Without loss of generality, we choose $\textbf{F}_i(\textbf{H}_i,r^*)$ so that the herders, in the absence of nearby targets, converge towards the origin if outside the goal region of radius $r^*$; namely we set
\begin{equation}
    \textbf{F}_i(\textbf{H}_i,r^*)=\left\{\begin{aligned}
        -&v_H\widehat{\textbf{H}}_i \quad &\text{if $|\textbf{H}_i|\geq r^*$}\\
        &0 \quad &\text{otherwise}
    \end{aligned}\right . 
\end{equation}

As typically done in the control theoretic and robotics literature, e.g. \cite{auletta2022herding}, we assume that at each time step, herder $i$ selects a target within its sensing region, say $\textbf{T}_i^*=\textbf{T}_i(\textbf{H},\textbf{T},\xi)$, to coral and chase. Then, we choose 
\begin{equation}
   \textbf{I}_i (\textbf{T},\textbf{H},\xi)= -\left[\alpha \left( \textbf{H}_i-(\textbf{T}^*_i +\delta  \widehat{\textbf{T}}^*_i)\right )   \right]_{v_H}
   \label{eqn:feedback}
\end{equation}
where 
$\delta=\lambda/2$ is the distance at which the herder places itself behind the chosen target to coral it towards the goal region, $\alpha$ is a positive dimensional constant and $[\cdot]_{v_H}$ 
is a saturation operator that limits the herders' maximum speed to $v_H$ when chasing a target.

The target to chase $\textbf{T}^*_i$ is selected by herder $i$ 
as the target with the largest distance from the origin among those, if any, within the sensing radius of herder $i$. Furthermore, if herder $i$ detects other herders $\textbf{H}_j$ in its sensing region (i.e. such that $|\textbf{H}_j-\textbf{H}_i|\leq \xi$), it only considers those targets $\textbf{T}_a$ for which $|\textbf{T}_a -\textbf{H}_i|\leq|\textbf{T}_a -\textbf{H}_j|$. 
Through this simple local rule, nearby herders effectively cooperate so as to decide which target to chase without needing any global information on the positions of other herders and targets.

We assume that the herders' velocity $v_H>v_T$ as typically done in the control literature \cite{long2020comprehensive} and to prevent the formation of stable chasing patterns that can be observed for   $v_T\lesssim v_H$  \cite{you2020nonreciprocity,fruchart2021non,saha2020scalar}, and that can hinder the achievement of the control goal. In addition, the radius of the repulsion zone, $\lambda$, is chosen smaller than that of the sensing area, $\xi$, as any other choice would be unrealistic. 


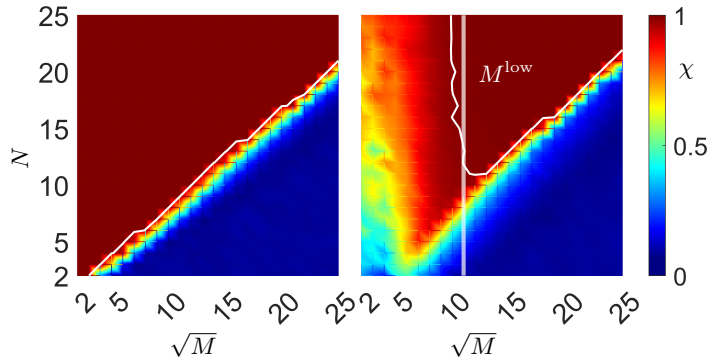
\begin{figure}[t]
\centering
\begin{overpic}[scale=0.52]{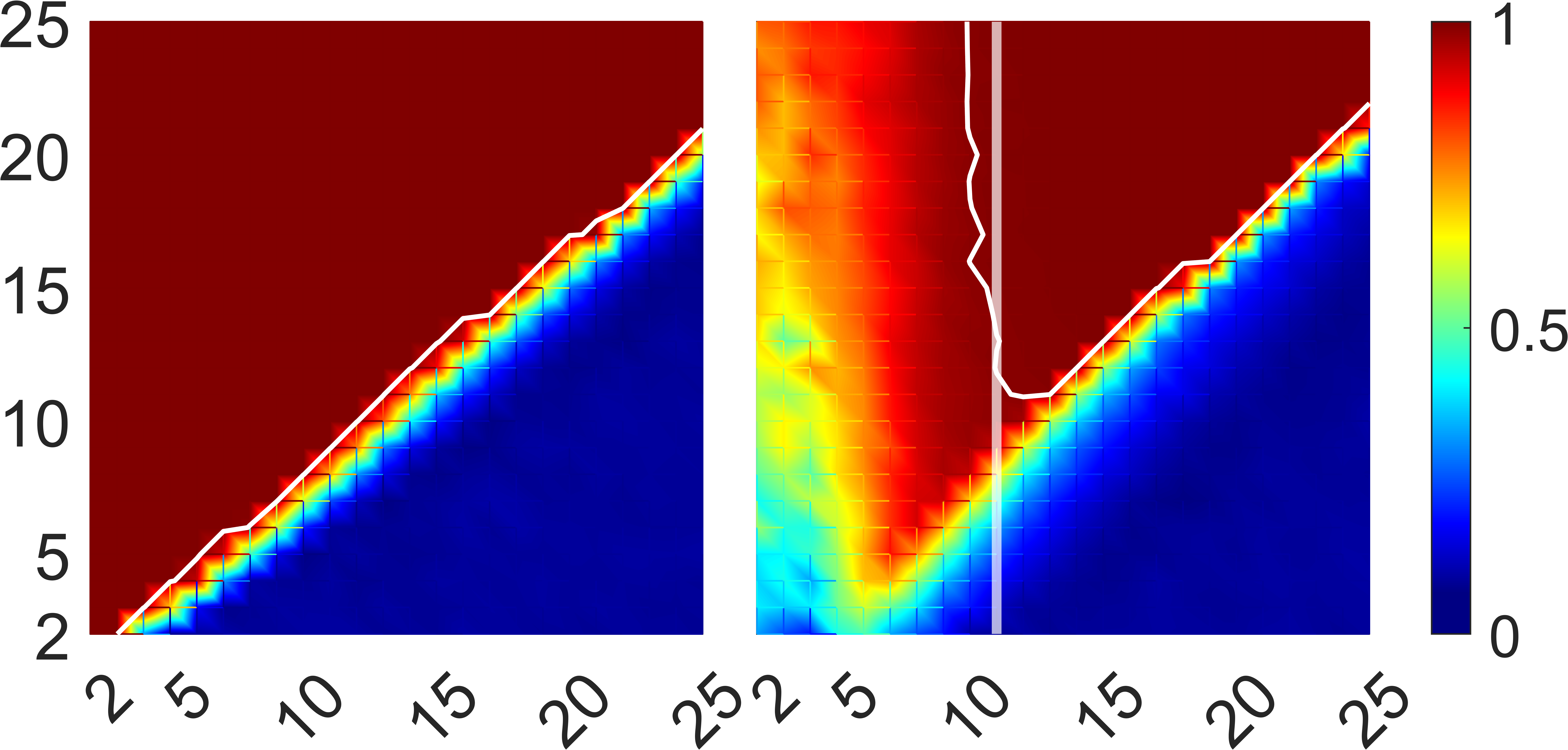}
\put(23,-3){\makebox(0,0){\textbf{$\sqrt{M}$}}}
\put(65,-3){\makebox(0,0){\textbf{$\sqrt{M}$}}}
\put(-3,25){\makebox(0,0){\rotatebox{90}{\textbf{$N$}}}}
\put(97,38){\makebox(0,0){\textbf{$\chi$}}}
\put(70,38){\makebox(0,0){\textcolor{white}{$\mlow$}}}
\end{overpic}
 \\ \vspace{0.5cm}
\caption{ \label{fig:chi} Values of the fraction $\chi$ of successfully herd targets obtained for different values of $M$ and $N$ when $R=50$. Results are averaged over $50$ simulations; the increments of $N$ and $\sqrt{M}$ have values $\Delta N=1$, $\Delta \sqrt{M}=1$. The level curve for $\chi^*=0.99$ is depicted by the white curve. The left panel shows the case of infinite sensing ($\xi=\infty$) where $N^*\propto\sqrt{M}$ while the right panel the case of limited sensing ($\xi<\infty$) where  we recover $N^*\propto\sqrt{M}$ only above a critical threshold $M>\mlow$ (white vertical line).}
\end{figure}

Next, we study the {\em herdability} of a group of $M$  targets by a group of $N$ herders \footnote{This concept is related but not entirely equivalent to the definition of herdability of complex systems that was given independently in \cite{ruf2019herdability}}. 

Specifically, we define a group of $M$ target agents as ``herdable'' by $N$ herders if the latter  can successfully guide at least a certain fraction $\chi>\chi^*$ of the former towards $\og$ within a finite time (see SM for further details). The threshold fraction $ \chi^*$ is set based on standard values in control theory, typically $\chi^* \in \{0.9, 0.95, 0.99\}$ \cite{aastrom2021feedback}. 
Given the dynamics of the agents, we will then look for the \textit{minimal} number of herders, denoted as $N^*(M)$, required to achieve herdability of $M$ targets.

For the sake of comparison, we start by considering herders with infinite sensing capabilities, setting $\xi=\infty$.
As shown in Figure \ref{fig:chi}a, for a broad range of target group sizes, the required number of herders, $N^*(M)$, exhibits a quadratic relationship with the number of targets. Conversely, in scenarios with finite sensing (Figure \ref{fig:chi}b), the scaling $N^*(M) \propto \sqrt{M}$ is observed, but only when the number of targets, $M$, exceeds a certain critical threshold, $\mlow$. Below this threshold, the task notably demands more herders, indicating, counterintuitively, that fewer targets do not necessarily ease the control task with herders' limited sensing abilities.

In general, the minimum number of herders, $N^*(M)$, needed to shepherd $M$ targets depends on two conditions: (i) the herders must collectively sense all targets, which are random independent walkers, and (ii) they must counterbalance the diffusion of the $M$ targets with the transport flow they induce. 
From a simple dimensional argument, as the $M$ targets are distributed in a {\em two}-dimensional circular domain  while the $N$ herders tend to arrange themselves on its {\em one}-dimensional boundary [see Fig.\ref{fig:PRL_HCP}(b) and supplementary videos], condition (ii) is satisfied for $N^*(M)\propto\sqrt{M}$ (as observed in Fig. \hyperref[fig:chi]{\ref{fig:chi}a}) while condition (i) is trivially satisfied when the herders possess infinite sensing ($\xi=\infty$).

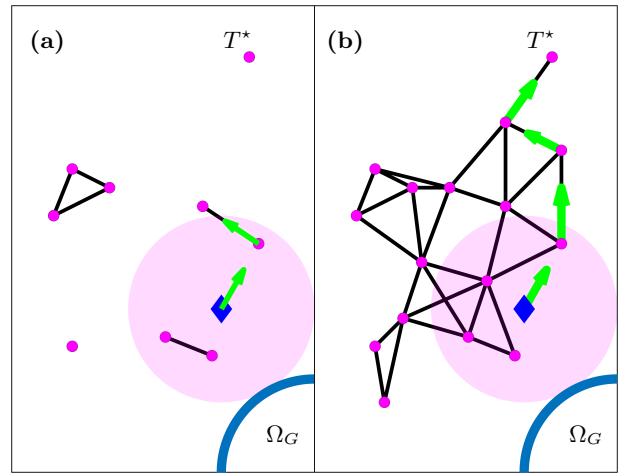
\begin{figure}
\centering  
\begin{overpic}[scale=0.7]{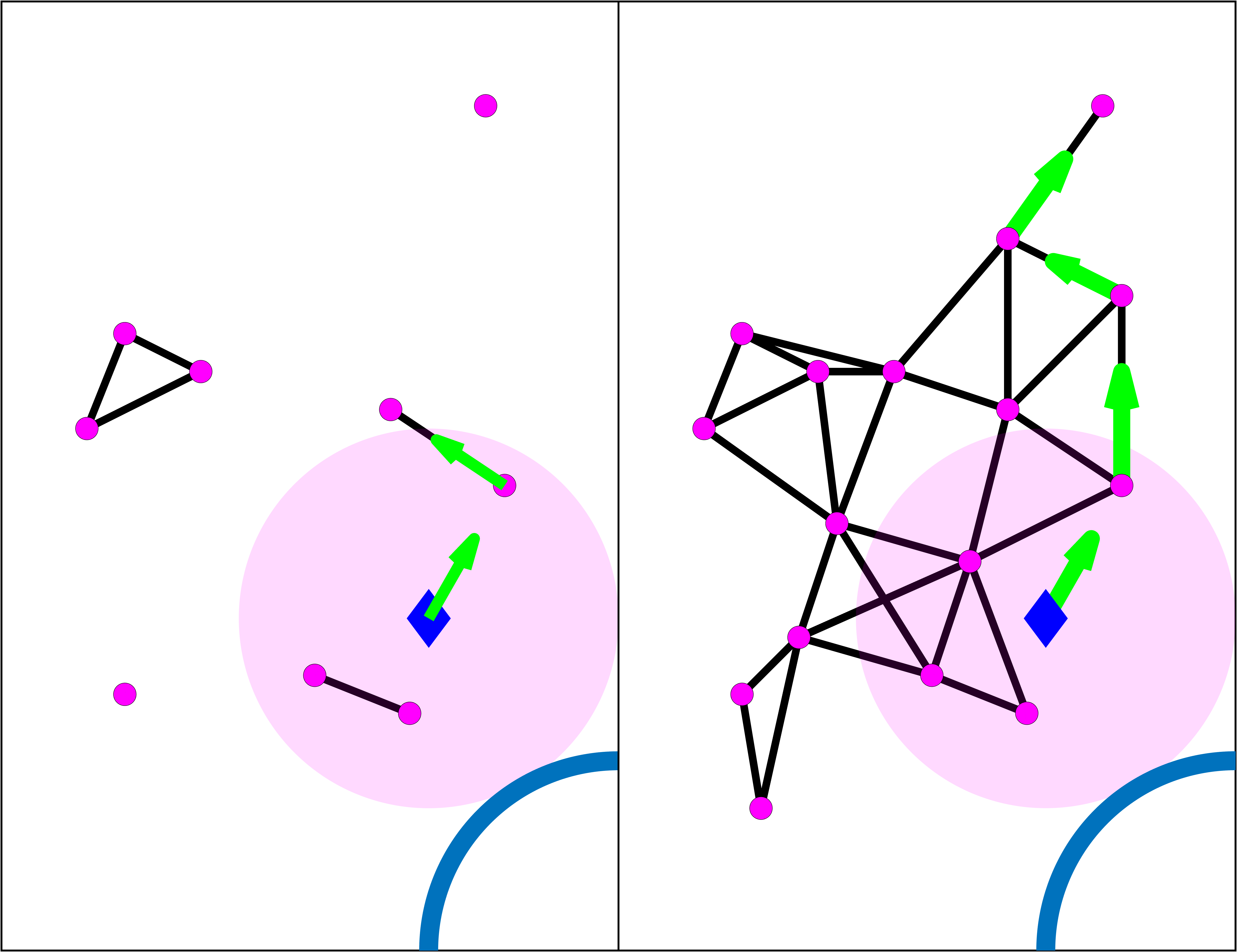}
\put(3,70 ){{\textbf{(a)}}}
\put(52,70 ){{\textbf{(b)}}}
\put(85,70 ){{\textbf{$T^\star$}}}
\put(35,70 ){{\textbf{$T^\star$}}}
\put(42,5 ){{\textbf{$\og$}}}
\put(92,5 ){{\textbf{$\og$}}}
\end{overpic}
\caption{Two representative configurations of targets and the structure of the corresponding herdability graph $\mathcal{G}$ (whose edges are depicted as solid black lines) (a) below and (b) above the critical percolation threshold $\widehat{\mlow}$. Green arrows show possible paths the herder could potentially navigate to reach the furthermost targets, denoted as $\textbf{T}^{\star}$, showing that when the graph is too sparse [panel (a)] more distant targets can be lost.}\label{fig:fig3idea}
\end{figure}

However, with finite sensing $\xi < \infty$, meeting condition (i)  becomes increasingly more cumbersome as target density decreases (e.g., $M < \mlow$).   
In this case, targets can become too sparse, hindering herders from efficiently scouting the area based on local information alone. Consequently, a larger number of herders, $N^*$ is required to ensure all targets, particularly those farthest from the goal $\og$, are observed consistently. This requirement deviates from the quadratic scaling observed with infinite sensing. For $M > \mlow$, the higher density of targets enables herders to effectively navigate end explore the area of interest moving from target to target, even without sensing each target at every time instant, thus aligning with the scaling law observed in the infinite sensing scenario.

To explain the critical threshold $\mlow$, we analyze how herders, relying on local information, can satisfy the condition of sensing and corralling also distant targets from $\og$.

To this aim, we define the herdability graph $\mathcal{G}$ as the random geometric graph \cite{BARTHELEMY20111} where nodes represent targets, and an edge exists between two targets, say $\textbf{T}_a$ and $\textbf{T}_b$, if their distance is within the sensing radius of the herders, i.e. if $\vert \textbf{T}_a - \textbf{T}_b \vert \leq \xi$.  

Then, a path in $\mathcal{G}$ from one target, $T_a$, to another generic target, say $T_c$, indicates the potential for a herder to transition from sensing $T_a$ to sensing $T_c$. Therefore, we propose to estimate the critical threshold $\mlow$ by calculating the percolation threshold of the graph $\mathcal{G}$, denoted as $\widehat{\mlow}(R,\xi)$; in particular, we compute $\widehat{\mlow}(R,\xi)$ in the worst-case at $t=0$  when targets are randomly and uniformly distributed within a circle of radius $R$ (See Section II of the SM for further details). 

Fig.\ref{fig:fig3idea} presents two schematic examples illustrating target configurations below and above the estimated threshold $\widehat{\mlow}$ along with their respective herdability graph structures $\mathcal{G}$. These examples clarify how $\widehat{\mlow}$ can serve as an approximation for the critical threshold $\mlow$.
\begin{figure}[tb]
    \centering
\begin{overpic}[scale=0.6]{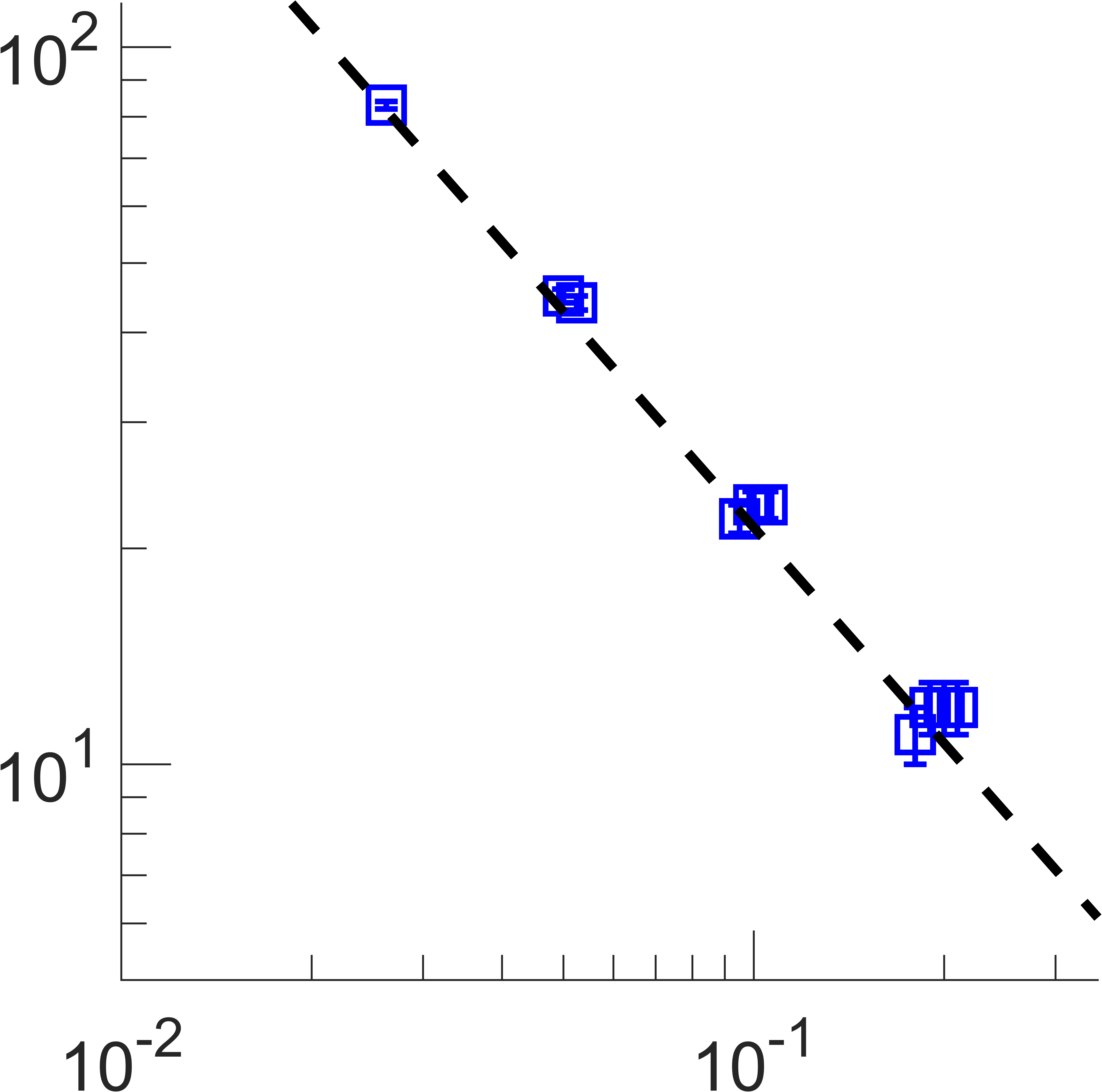}
\put(43,2){{\textbf{$\xi/R$}}}
\put(2,43){\rotatebox{90}{\textbf{$\sqrt{\mlow}$}}}

\end{overpic}
    \caption{Scaling of the critical threshold $\mlow$ as a function of $\xi/R$. The numerically observed values of $\mlow$ (scatter dots), evaluated by direct inspection, are compared with the theoretical estimate $\widehat{\mlow}$ (dashed line) for different values of $\xi$ and $R$ (see  Tab... in the SM for the values of $\xi$ and $R$ selected). Error bars represent the maximum precision of the computation given the stepsize $\Delta \sqrt{M}=1$ used in the simulations. Results for  $\chi^*\in\{0.90,\,0.95\}$ are reported in section I of the Supplementary Material confirming the observed scaling. For the same $\xi/R$ value, scatter points were shifted on the $x$-axis to increase visibility.}
    \label{fig:fig4idea}
\end{figure}
%
Given the known scaling of the percolation threshold of a two-dimensional geometric graph as $R^2/\xi^2$ \cite{dall2002random}, we anticipate $\mlow \sim R^2/\xi^2$. This expectation aligns with our numerical findings shown in Figure \ref{fig:fig4idea} when $\chi^* = 0.99$, confirming that our theoretical approach effectively captures the observed trend. For more details and additional validation for different $\chi^*$ values and noise levels in the targets dynamics, we refer the reader to the SM.

%
Our study's central finding is that effective herdability hinges on sufficient connectivity of the herdability graph $\mathcal{G}$. To analytically substantiate this, we examine a  simpler one-dimensional scenario and derive a PDE characterizing the spatio-temporal dynamics of the herders' density, denoted as $\rho^H$. A pivotal aspect of our analysis involves translating the decision-making process herders use to select the target to chase, $\mathbf{T}^*_{i}$, into a continuum framework. We propose this can be done by expressing the target selection rule used by the herders as a weighted average approximated by:
\begin{equation}
\textbf{T}^*_i  = \lim_{\gamma\to\infty}\dfrac{\sum _{a\in\mathcal{N}_i}e^{\gamma |\textbf{T}_a|}\,\textbf{T}_a }{\sum _{a\in\mathcal{N}_i}e^{\gamma |\textbf{T}_a|}}
    \label{eqn:continuum_selection}
\end{equation}
with $\mathcal{N}_i$ being the set of target indexes such that $\mathbf{d}_{ia}\leq \xi$.

Then, recasting (\ref{eqn:continuum_selection}) in a continuum framework, ignoring the herders' own dynamics, $\textbf{F}_i$, and setting $\delta=0$ in \eqref{eqn:feedback} for the sake of simplicity, we capture heuristically the dynamics of the herders' density $\rho^H(x,t)$ by the following PDE:
\begin{equation}
    \rho^H_t + \left[-\frac{d V}{d x} \rho^H \right]_x=0
    \label{eqn:PDE_dyn}
\end{equation}
where
\begin{equation}
    -\frac{d V}{d x}=-\frac{1}{\mathcal{M}}\int_{\mathbf{B}_{\xi}(x)}e^{\gamma |y|}\rho^T(y)(x-y)\,dy
    \label{eqn:PDE}
\end{equation}
with $\rho^T$ being the density of the targets supposed to be stationary when the herders are sufficiently faster than the targets, and $\mathbf{B}_{\xi}(x)$ denotes a ball of radius $\xi$ centered in $x$, and $\mathcal{M}$ is a normalization factor.

\begin{figure}[t]
    \centering
    \includegraphics[width=.4\textwidth]{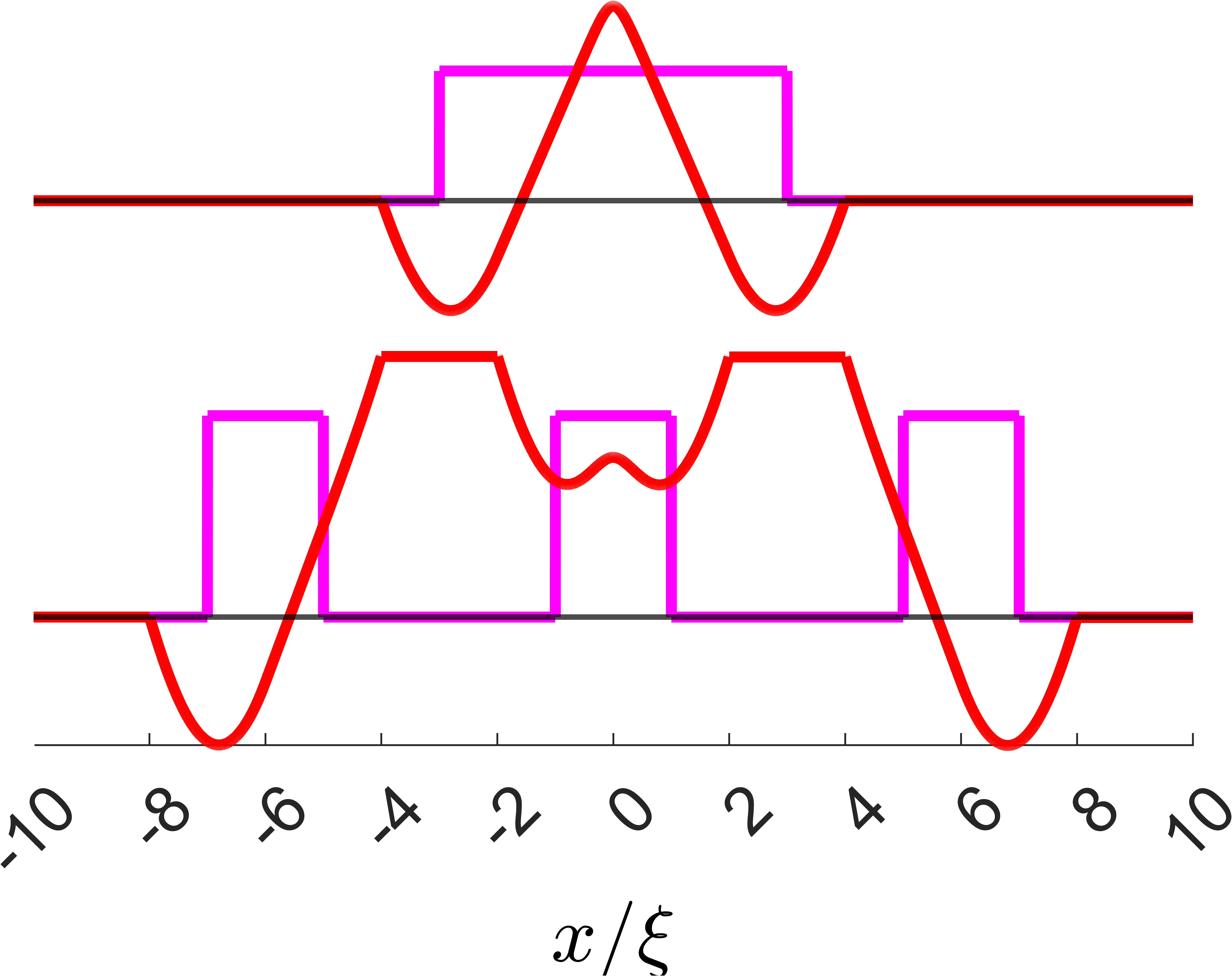}
    \caption{Stationary distributions of the targets, $\rho^T$, over a one-dimensional domain together with the corresponding potential $V$ (red line) computed by \eqref{eqn:PDE} showing global stability when no regions exist where $\rho^T|_{x\in\Delta}=0$ with $|\Delta| > \xi$ (top panel) and local stability otherwise (bottom panel). See supplementary video 3 in the SM for a simulation of \eqref{eqn:PDE_dyn} in the above scenarios.}  
    \label{fig:PDE}
\end{figure}

Using \eqref{eqn:PDE_dyn}, we find, as shown in Fig. \ref{fig:PDE}, that
a globally stable equilibrium configuration, in which herders completely encircle all targets, is attainable only under the condition that no regions exist where $\rho^T|_{x\in\Delta}=0$ with $|\Delta| > \xi$; a situation corresponding to the herdability graph $\mathcal{G}$ being disconnected. This further substantiates our finding that the targets need to maintain a sufficient level of connectivity within the herdability graph $\mathcal{G}$ in order to enable the herders to collectively detect and guide even the most distant ones.


In summary, the relevance of our findings is twofold. Firstly, we solved the shepherding control problem  assuming neither cohesive behaviour in the targets nor unlimited sensing capabilities of the herders. Secondly, we introduced and analysed the {\em herdability} of a group of targets by a group of herders providing conditions for the herders to coral and chase all targets. This is achieved by means solely of local information and knowledge of the control goal enabling the herders to select the next target to chase according to a decentralized feedback control mechanism. We found that the minimal number of herders $N^*$ required to successfully shepherd a group of $M$ targets scales as $N^*\propto \sqrt{M}$ only above a critical threshold $\mlow$ in the case of limited sensing. We showed that an estimate of such critical threshold can be effectively obtained in terms of the percolation of a suitably defined {\em herdability graph}.
Finally, we provided a strategy to translate to a continuum framework the target selection strategy used by the herders and, consequently, derived an appropriate PDE describing the evolution of the herders' density in a one-dimensional representative setting. The stability analysis of the asymptotic herders' distributions confirmed our claim that in order for the task to be successful the herdability graph must be adequately connected.

Ongoing and future work will be devoted to further explore the  continuum formulation of the shepherding control problem. Indeed, we anticipate that such a formulation will allow to investigate herdability for a variety of targets dynamics, leveraging on results from the literature on the physics of non-reciprocal interactions \cite{fruchart2021non} and on the control of PDEs, e.g., \cite{zuazua1990uniform,krstic2008boundary}. For example, the collective dynamics of the targets could be described as a wave equation where perturbations can effectively propagate in space, e.g. \cite{attanasi2014information}, and then solutions to the boundary-control problem of the wave equation described in \cite{zuazua1990uniform} could be adapted to find novel herder configurations to solve the shepherding control problem.

All the numerical simulations where carried out in MATLAB using a forward Euler scheme for the herders, and a Euler–Maruyama scheme for the targets, with time steps $\Delta t=0.03$, and total duration $t$ depending on the settling time required for stationarity (see Fig. S8 of the Supplementary Material). Parameters of targets' dynamics where chosen as $D=1$, $\beta=3$, $\lambda=2.5$. Parameters of herders' dynamics where chosen as $v_H=8$, $\alpha=5$, $\gamma=5$. The radius $r^*$ of the goal region $\og$ was chosen as $R/5$.

\bibliography{biblio}

\end{document}


\preprint{APS/123-QED}

\title{SUPPLEMENTARY MATERIAL\\
for\\
"Shepherding control and herdability in complex multiagent systems"}

\author{Andrea Lama}
\email{andrea.lama-ssm@unina.it}
\affiliation{Scuola Superiore Meridionale, Naples, Italy}

\author{Mario di Bernardo}%
 \email{mario.dibernardo@unina.it}
\affiliation{Department of Electrical Engineering and ICT, University of Naples Federico II, Italy
}
\altaffiliation[Also at ]{Scuola Superiore Meridionale, Naples, Italy.}

\date{\today}
\begin{abstract}

\end{abstract}

\maketitle
\renewcommand{\thefigure}{S\arabic{figure}}
\renewcommand{\theHfigure}{S\arabic{figure}}

\section{Scaling analysis}
\label{mysection:scaling}

\begin{figure*}[h]
\begin{overpic}[width=.5\textwidth]{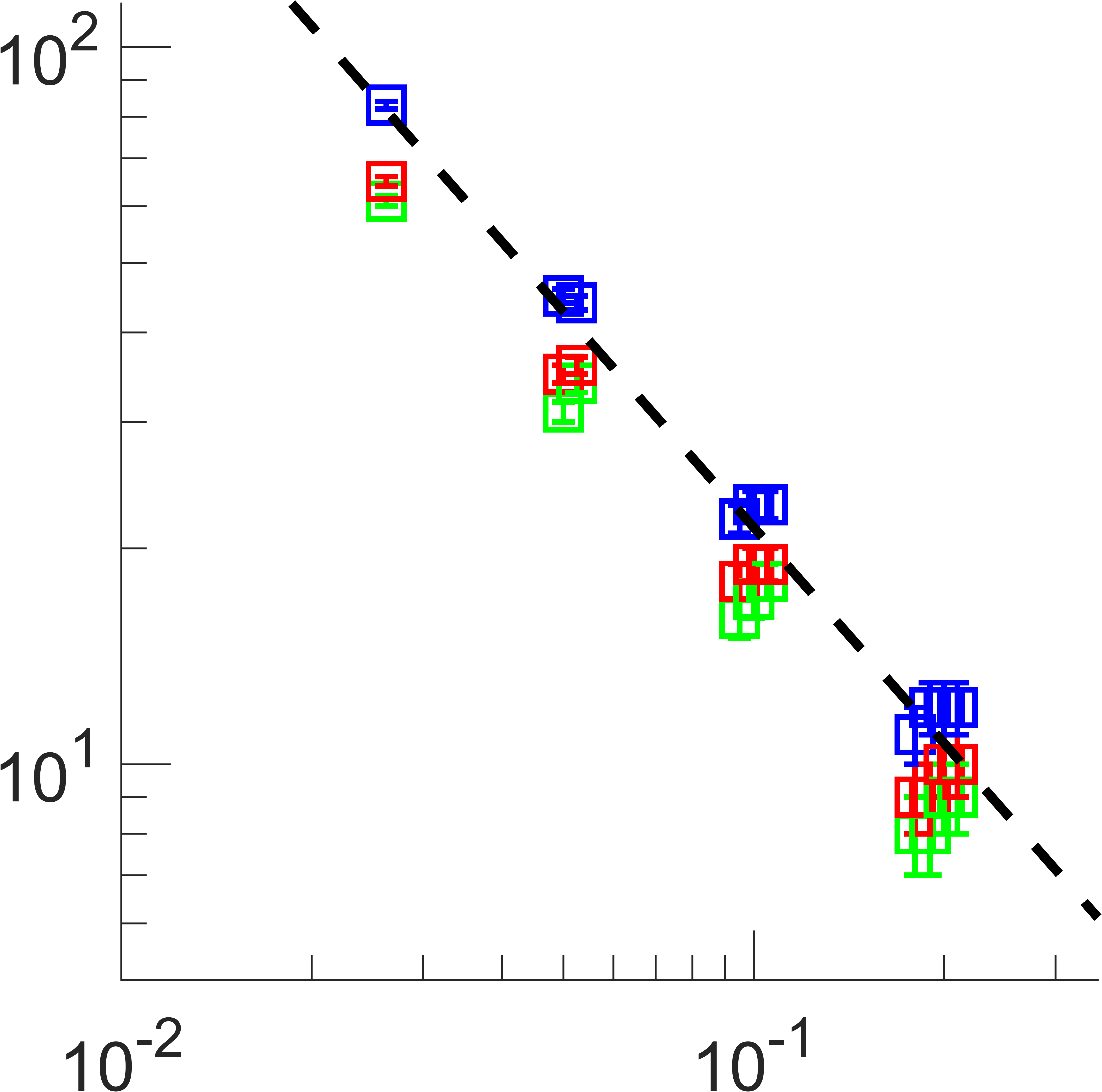}
  \put(50,2){\makebox(0,0){$\xi/R$}} 
  \put(2,50){\makebox(0,0){\rotatebox{90}{$\sqrt{\mlow}$,\  $\sqrt{\widehat{\mlow}}$}}} 
\end{overpic}
\caption{\label{fig:scaling_supp} Scaling of the critical threshold $\mlow$ as a function of the ratio between the sensing radius $\xi$ and the size of the initial region $R$. The numerical values of $\mlow$ are reported for three values of $\chi^*\in\{.90,\,.95,\,.99\}$ (green, red, and blue scatter dots respectively), and are compared with the percolation threshold estimate of the herdability graph $\widehat{\mlow}$ (dashed line) from Eq.\eqref{eqn:percolation}. The numerical values of $\mlow$ get closer and closer to the percolation threshold estimate as the value of $\chi^*$ is increased from $0.9$ to $0.99$. In all cases, the expected scaling of $\mlow$ is preserved. For the same $(\xi/R)$ value, scatter points were shifted on the $x$-axis to increase visibility.}
\end{figure*}

Fig. \ref{fig:scaling_supp} shows the scaling of the critical threshold $\mlow$, for three values of $\chi^*\in\{0.90,\,0.95,\,0.99\}$ as a function of $\xi/R$, where the numerically observed values of $\mlow$, evaluated by direct inspection, are compared with the percolation threshold estimate $\widehat{\mlow}$ of the herdability graph $\mathcal{G}$, obtained numerically as described in Section \ref{mysection:percolation}.
The numerically detected values of $\mlow$ become increasingly closer to the percolation threshold estimate as the value of $\chi^*$ is increased from $0.9$ to $0.99$. Most notably, in all cases the expected scaling of $\mlow$ with $\xi/R$ is preserved.

In Table \ref{tab:table1}, all the values are reported of $R$, $\xi$, and $\mlow$ for the three possible values of $\chi^*$.

\subsection{Noise effects}

\begin{figure}[h]
   \centering
    \begin{overpic}[width=.45\textwidth]{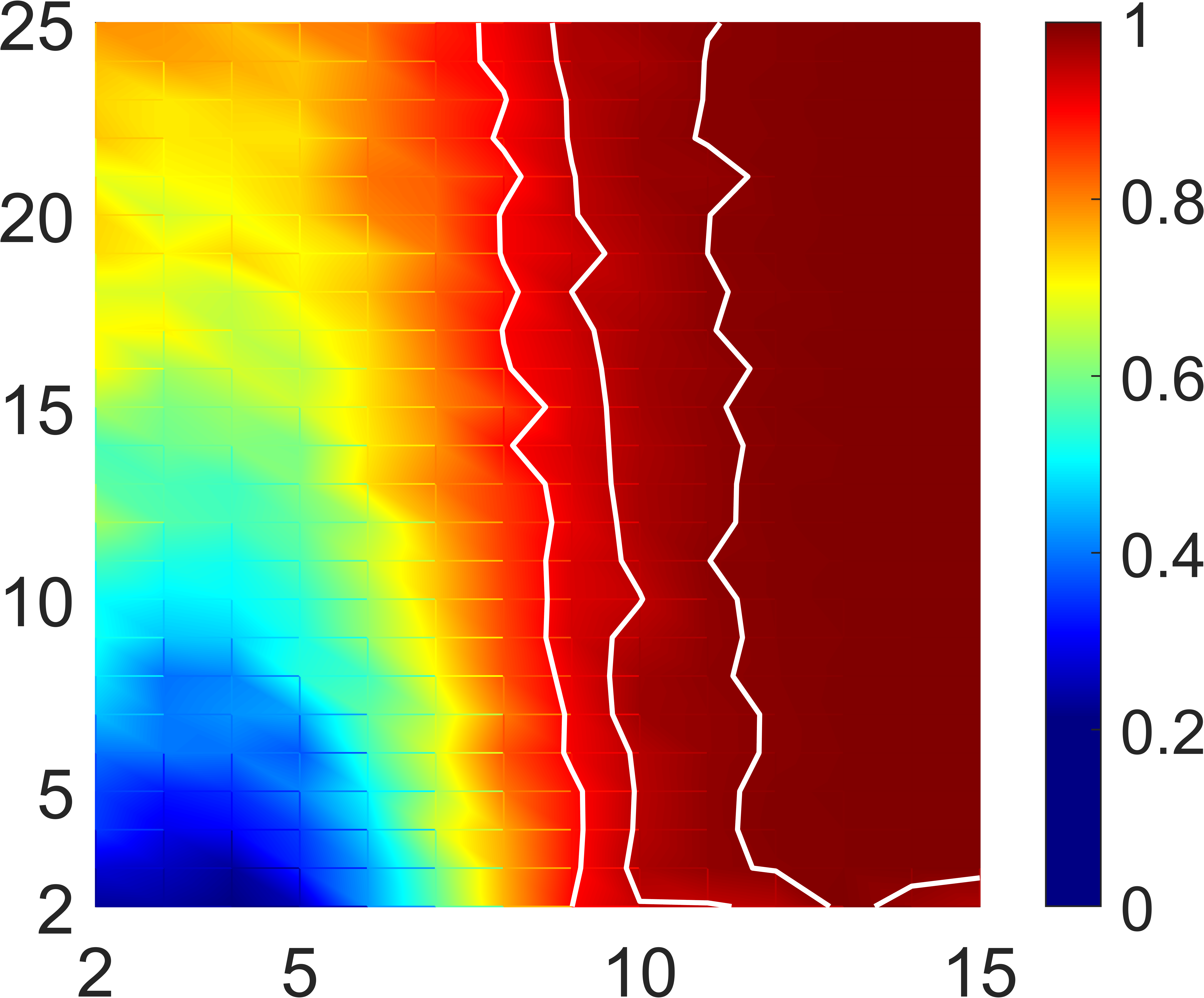}
        \put(-10,85){(a)}
        \put(-6,55){$N$}
        \put(35,0){$\sqrt{M}$}
         \put(100,58){$\chi$}
    \end{overpic}\vspace{.5cm}\\
    \begin{overpic}[width=.45\textwidth]{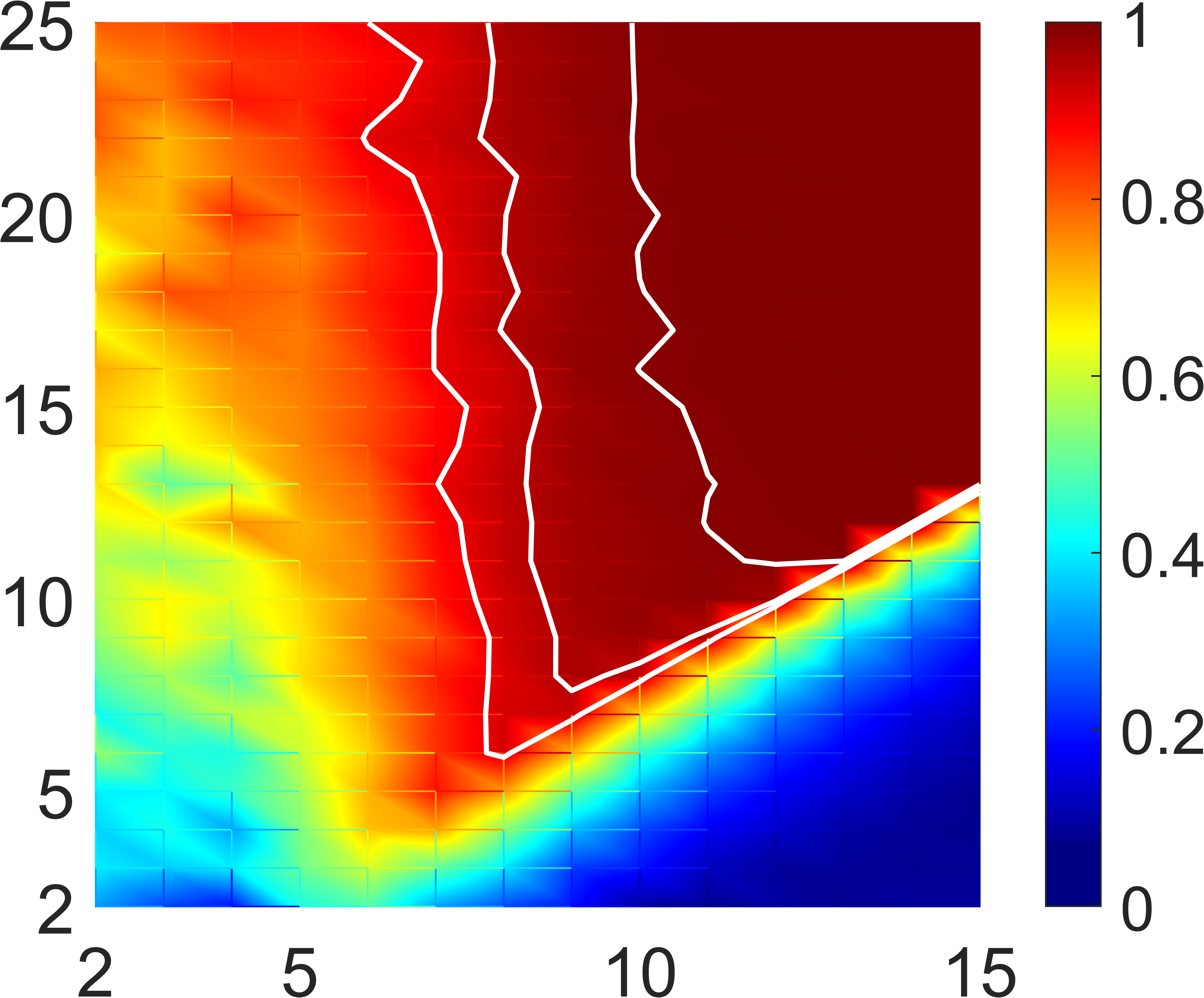}
        \put(-10,85){(b)}
        \put(-6,55){$N$}
        \put(35,0){$\sqrt{M}$}
        \put(100,58){$\chi$}
    \end{overpic}\vspace{.5cm}\\
    \begin{overpic}[width=.45\textwidth]{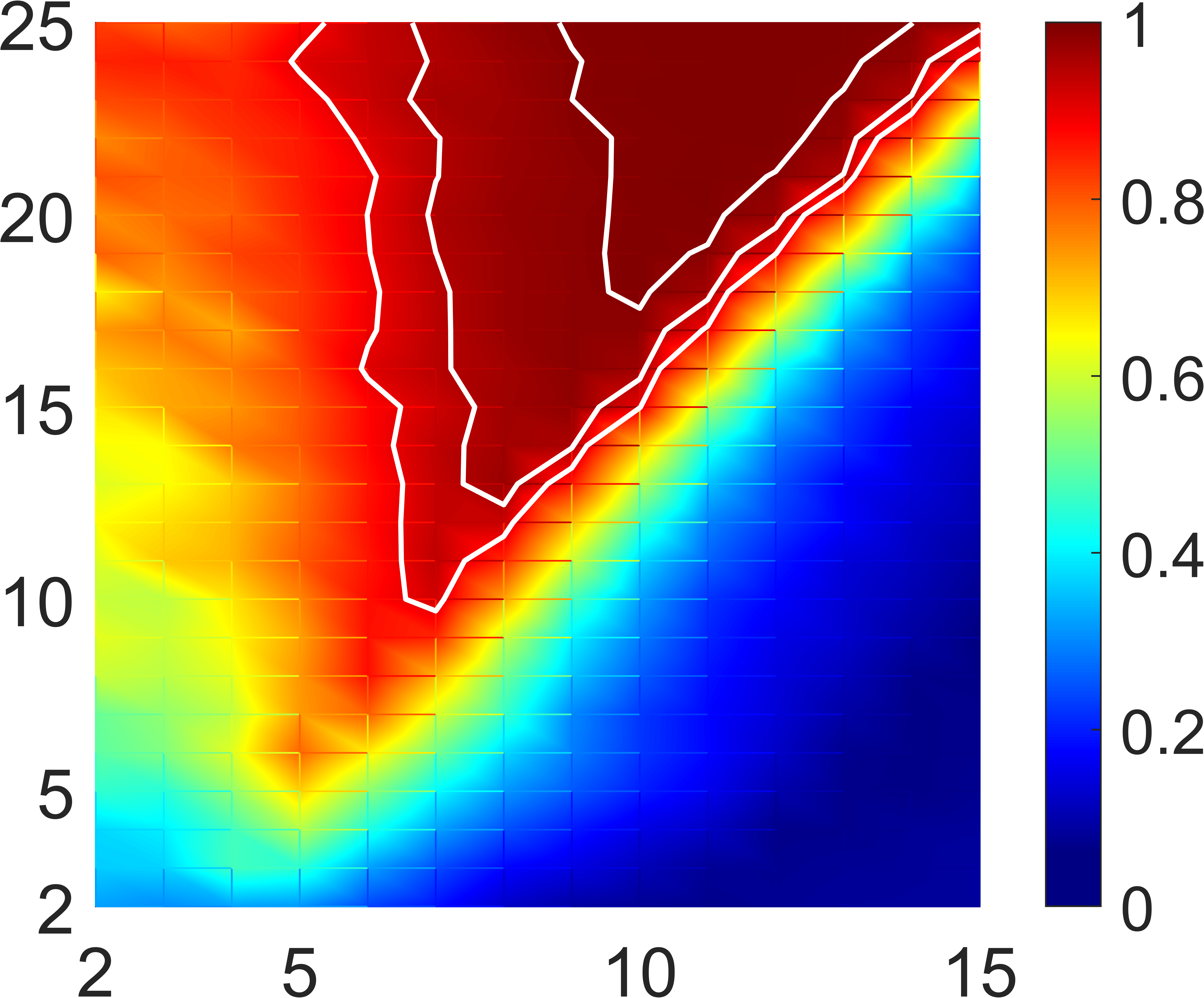}
        \put(-10,85){(c)}
        \put(-6,55){$N$}
        \put(35,0){$\sqrt{M}$}
        \put(100,58){$\chi$}
    \end{overpic}
    \caption{Comparison of the results for the fraction of successully herd targets $\chi$ at $R=50$ and $\xi=10$ for different noise levels, with the white lines indicating the success levels $\chi^*\in\{0.90,\,0.95,\,0.99\}$. In particular, here we show the cases (a) $D=10^{-2}$, (b) $D=1$ (the case discussed in the main text) and (c) $D=2$. We can clearly see that increasing the noise in targets' dynamics corresponds to a decrease in the observed value of the critical threshold $\mlow$. In particular, in (a) we notice that for low enough values of the noise in the dynamics of the targets, the prediction of the herdability graph fully captures the herdability of the targets. The results are averaged over $50$ simulations.}
    \label{fig:different_noise_50_10}
\end{figure}

We investigated the impact of varying noise levels by changing the parameter $D$ in the dynamics of the targets described in the main text; we kept all the other parameters in the dynamics unchanged and we only considered values of $D$ that preserved the hypothesis on the relative speeds of herders and targets discussed in the main text. In particular, we considered the cases $D=10^{-2}$ (Fig. \ref{fig:different_noise_50_10}a), $D=1$ (Fig. \ref{fig:different_noise_50_10}b, that is the case discussed in the main text), and $D=2$ (Fig. \ref{fig:different_noise_50_10}c). Looking at Fig. \ref{fig:different_noise_50_10}, we can immediately observe that increasing the value of $D$ in the targets' dynamics corresponds to a decrease of the observed value of the critical threshold $\mlow$. This reduction can be heuristically understood by considering that increased fluctuations in the targets' behavior result in more frequent formation and dissolution of links within the herdability graph, despite the graph's average connectivity remaining unchanged. These transient connections facilitate the herders, who are significantly faster than the targets, in their pursuit and corralling of the targets.

\begin{figure*}[h]
    \begin{overpic}[width=.5\textwidth]{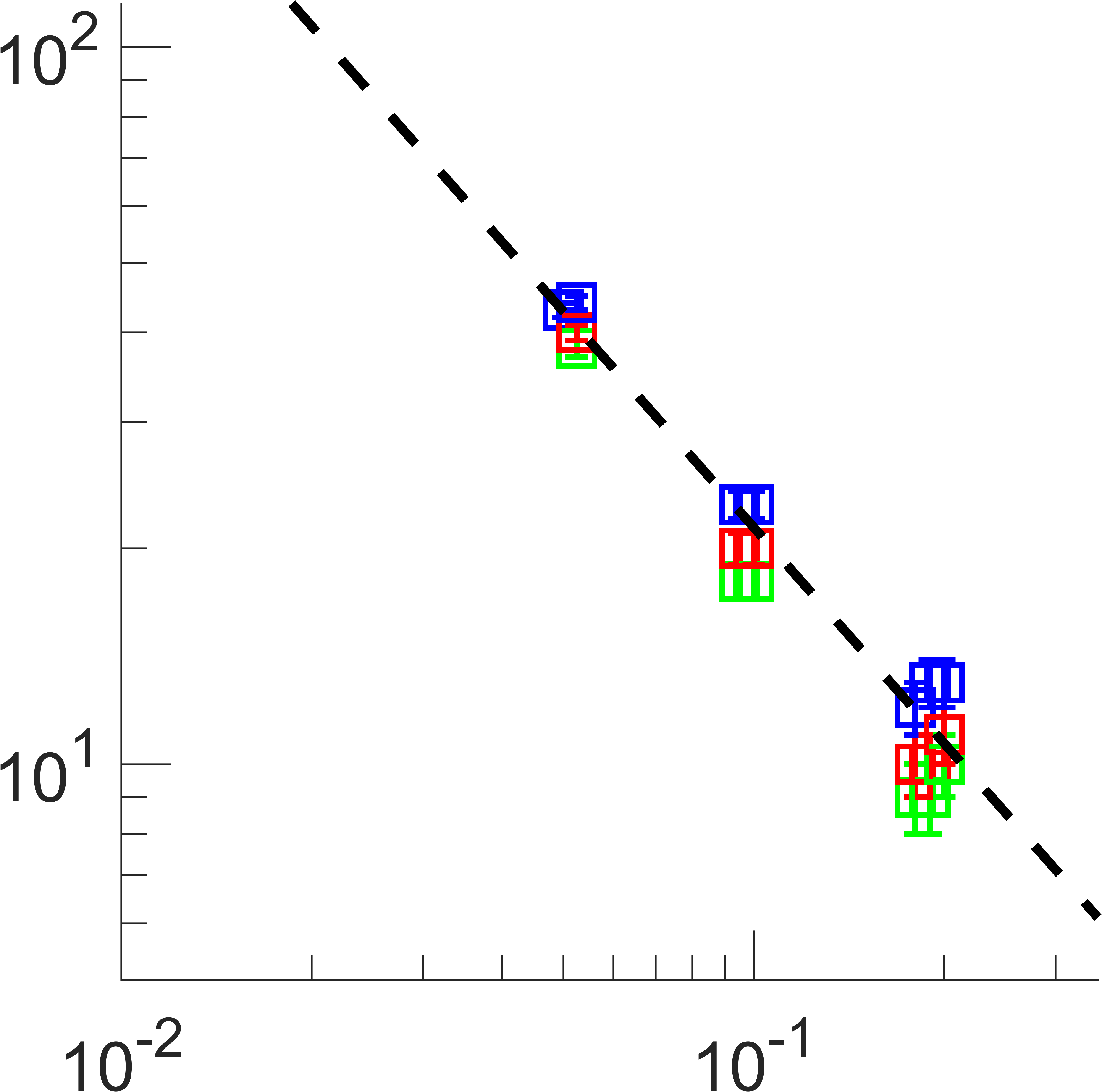}
    \put(50,2){\makebox(0,0){$\xi/R$}} 
  \put(2,50){\makebox(0,0){\rotatebox{90}{$\sqrt{\mlow}$,\  $\sqrt{\widehat{\mlow}}$}}} 
    \end{overpic}
\caption{\label{fig:scaling_supp_low_noise} Scaling of the critical threshold $\mlow$ as a function of the ratio between the sensing radius $\xi$ and the size of the initial region $R$ in the case $D=10^{-2}$. The numerical values of $\mlow$ are reported for for three values of $\chi^*\in\{.90,\,.95,\,.99\}$ (green, red, and blue scatter dots respectively), and are compared with the percolation threshold estimate of the herdability graph $\widehat{\mlow}$ (dashed line)   from Eq. \eqref{eqn:percolation}. For the same $(\xi/R)$ value, scatter points were shifted on the $x$-axis so to increase visibility.}
\end{figure*}

\begin{figure*}[h]
    \begin{tabular}{llll}
$R=50$ & \includegraphics[width=.2\linewidth,valign=m]{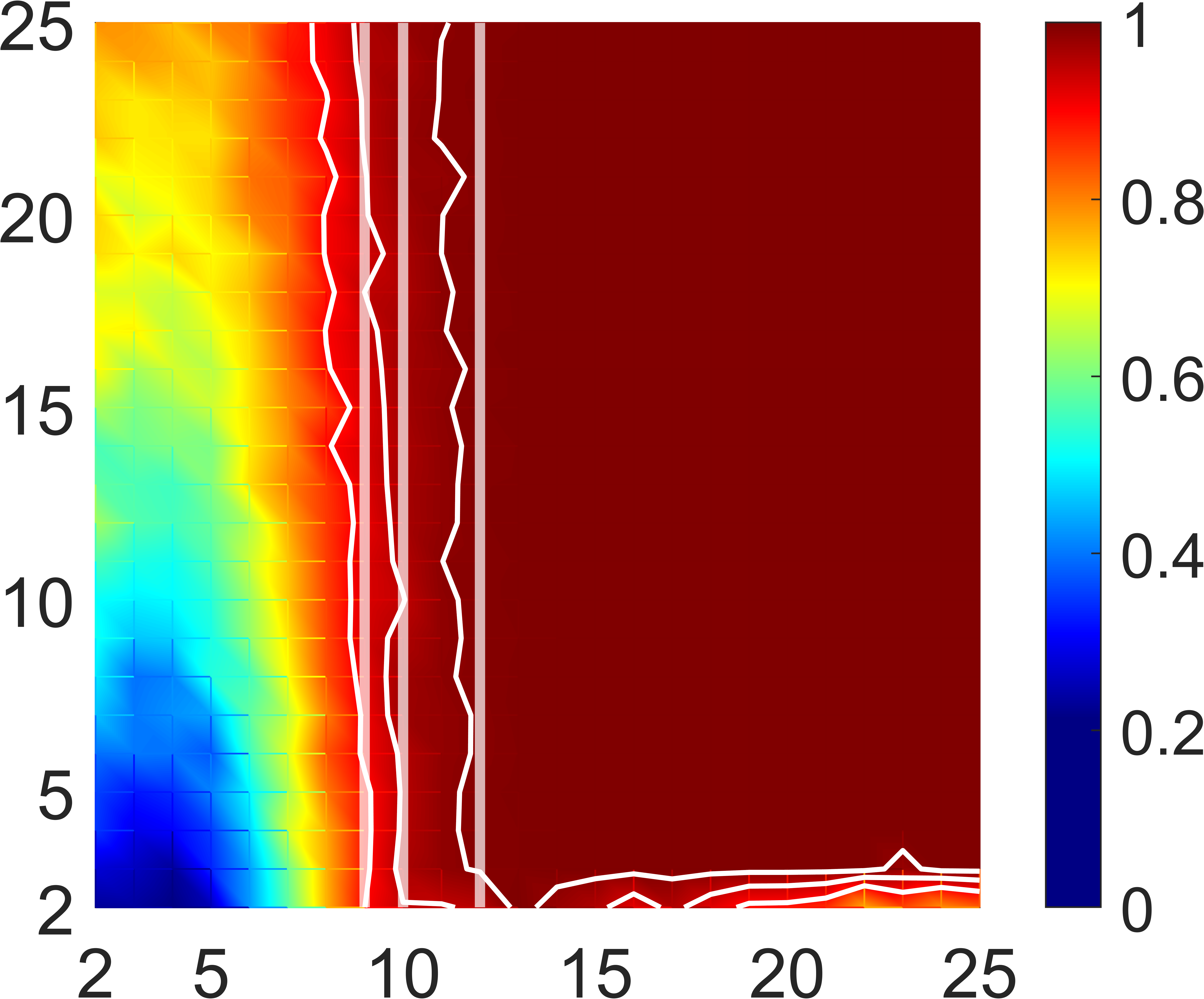} & &   \\
$R=100$ & \includegraphics[width=.2\linewidth,valign=m]{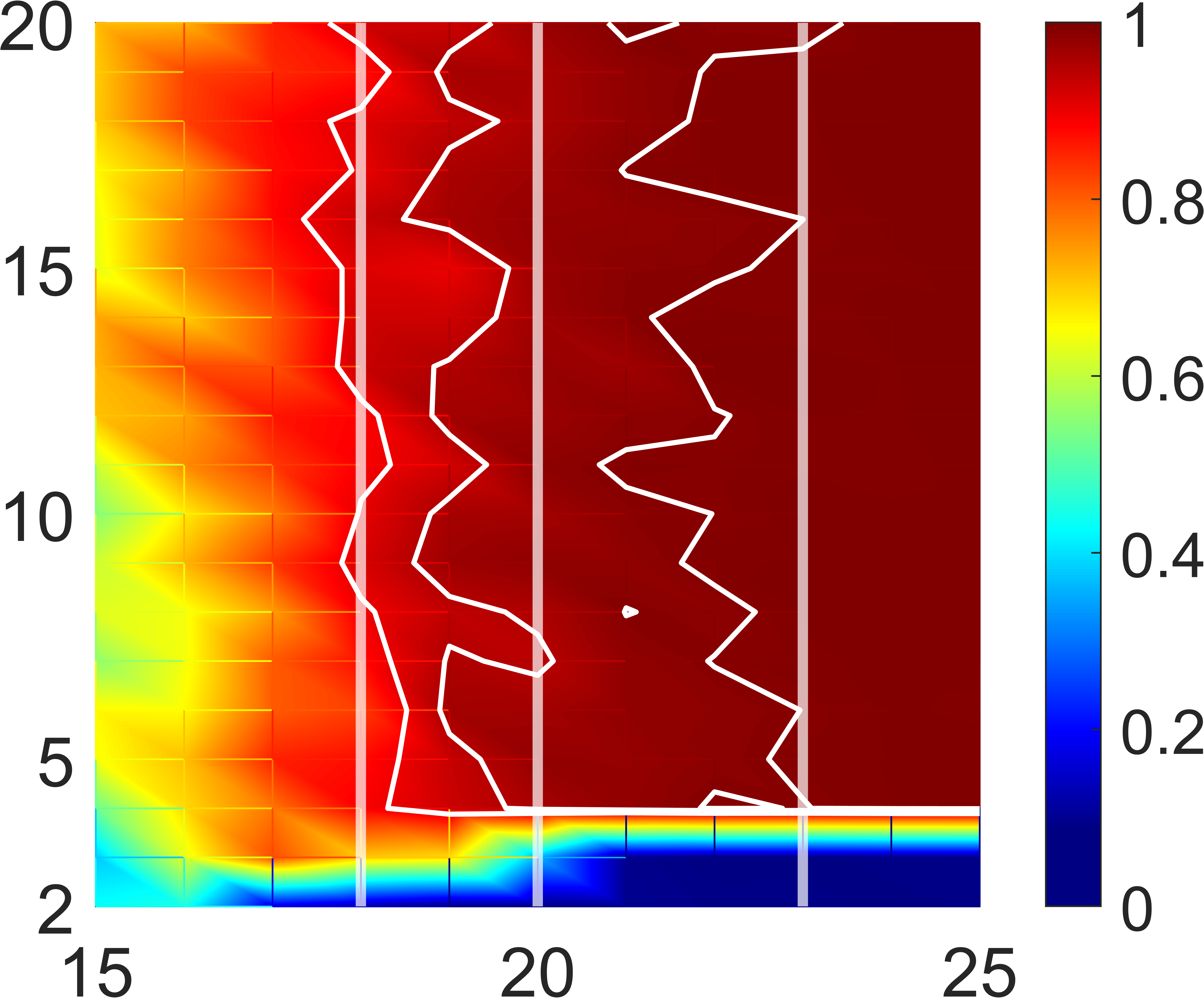}  & \includegraphics[width=.2\linewidth,valign=m]{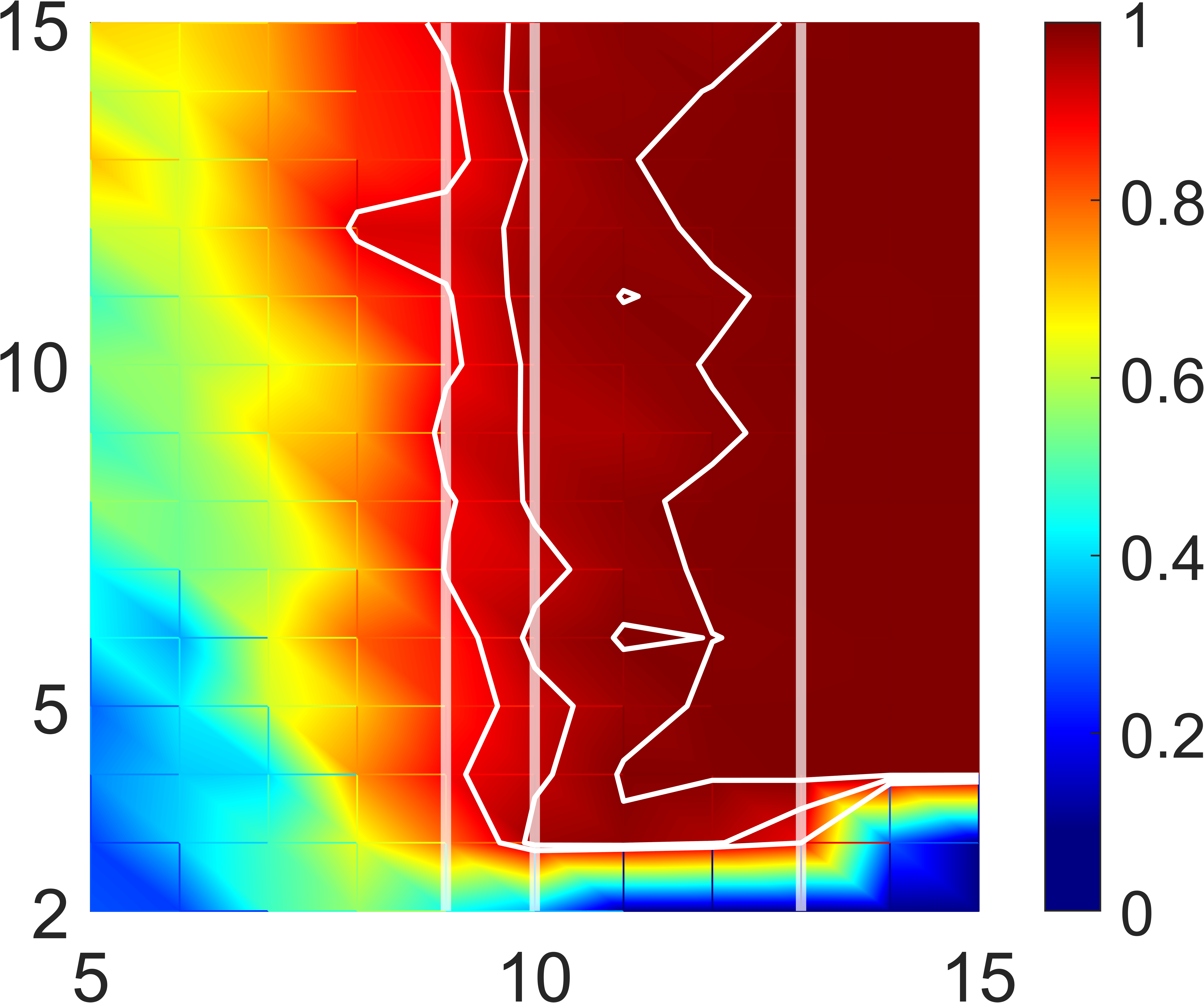} &    \\
$R=200$ & \includegraphics[width=.2\linewidth,valign=m]{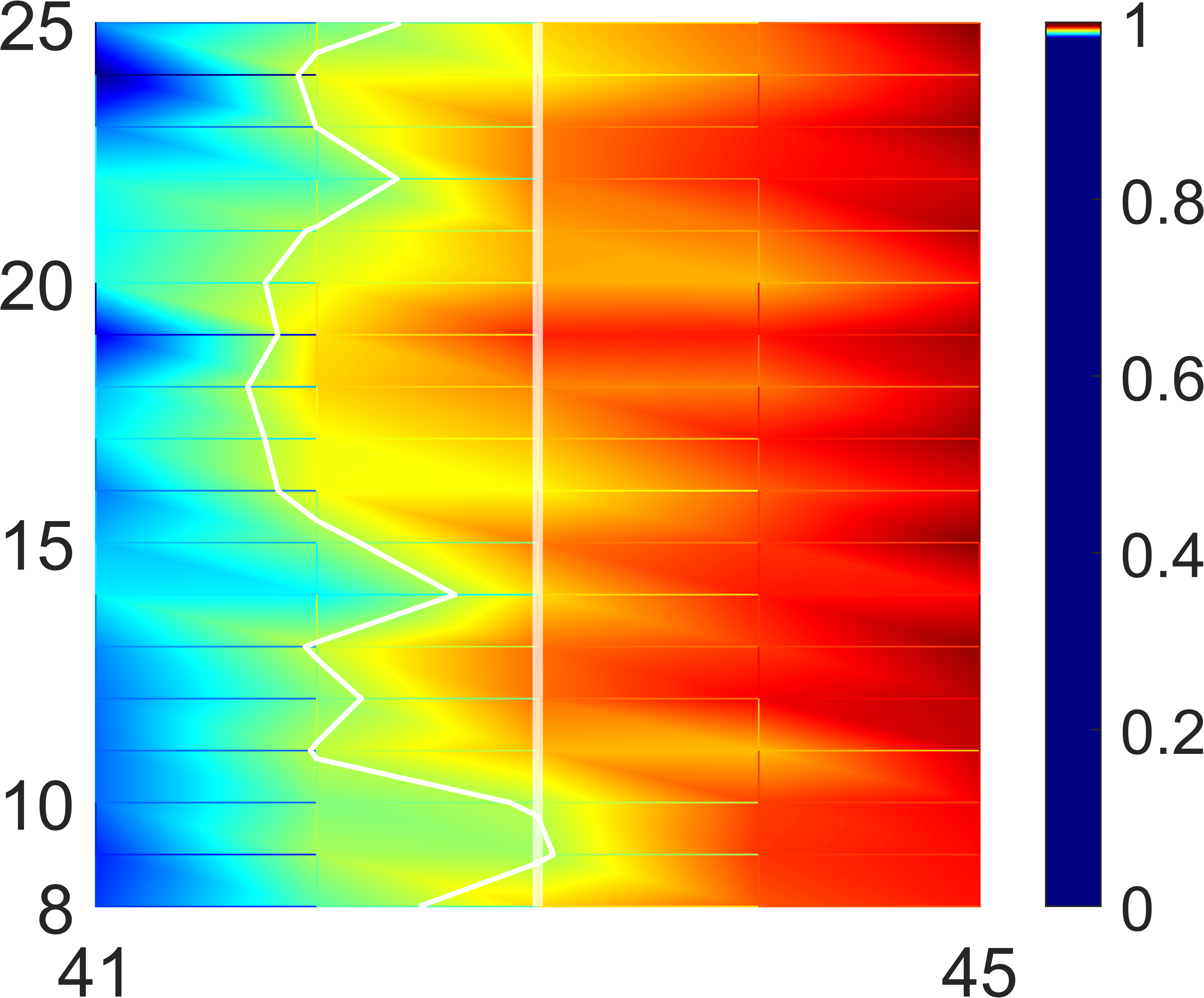}  & \includegraphics[width=.2\linewidth,valign=m]{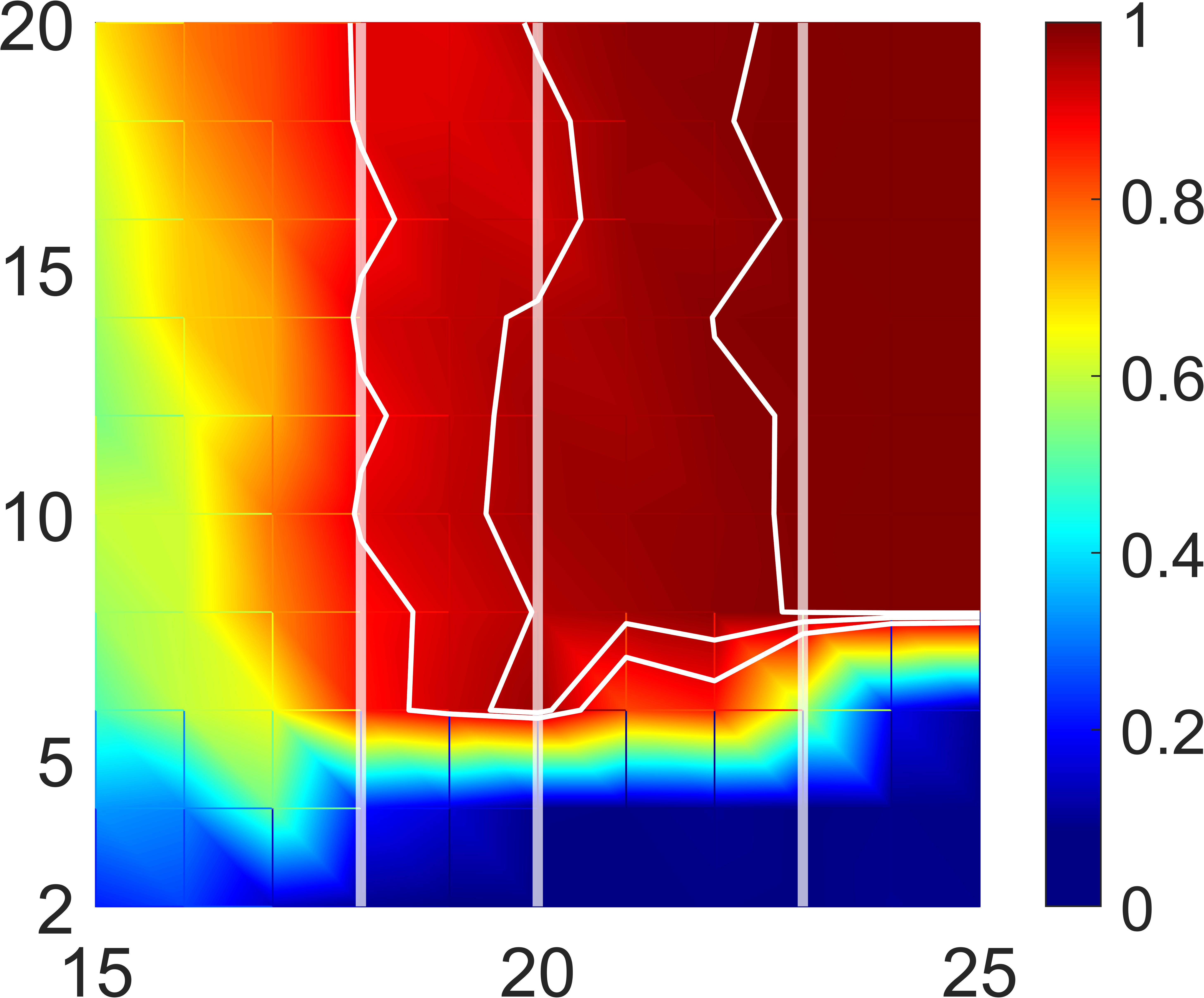}  & \includegraphics[width=.2\linewidth,valign=m]{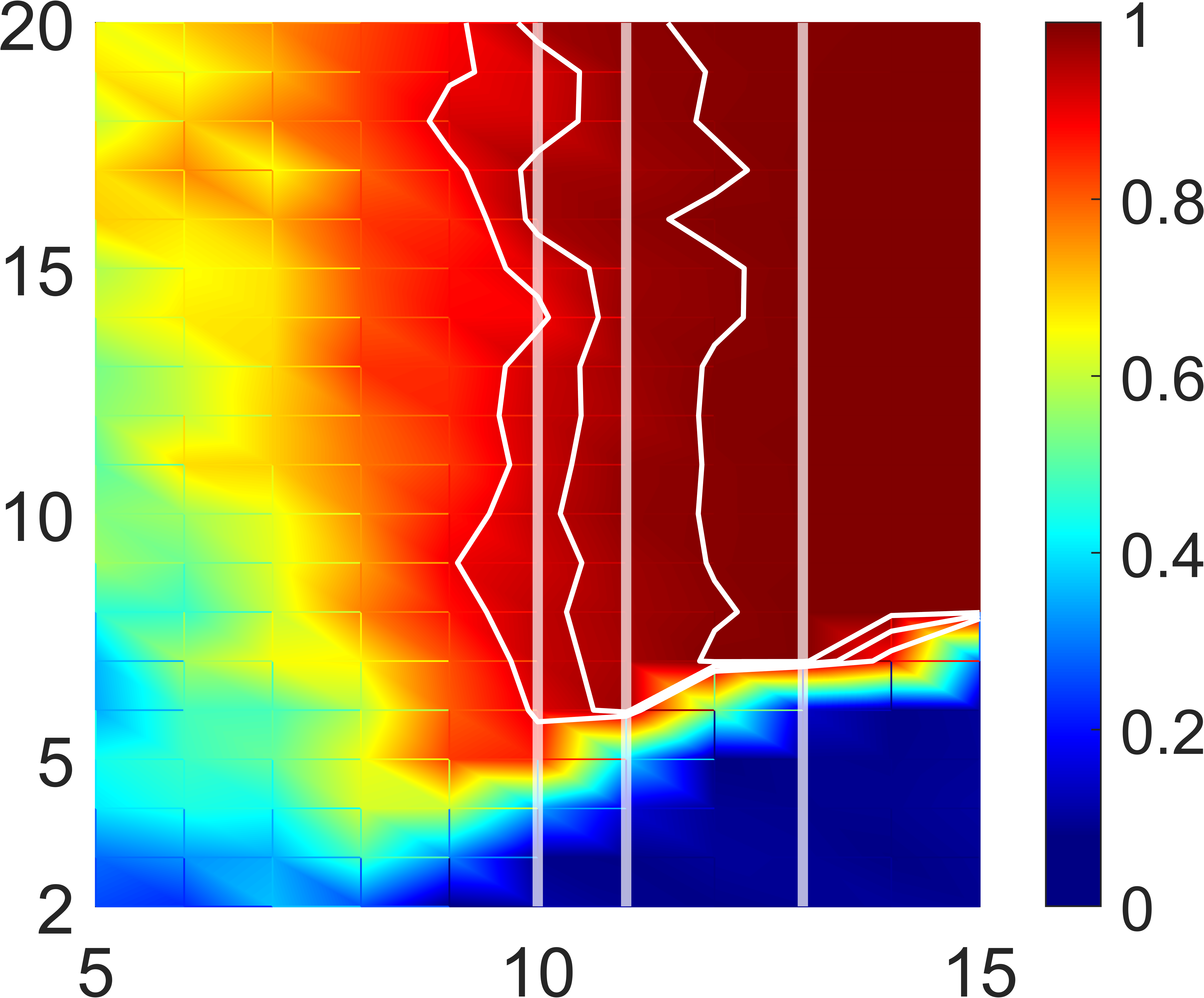} \\
$R=400$ &  & \includegraphics[width=.2\linewidth,valign=m]{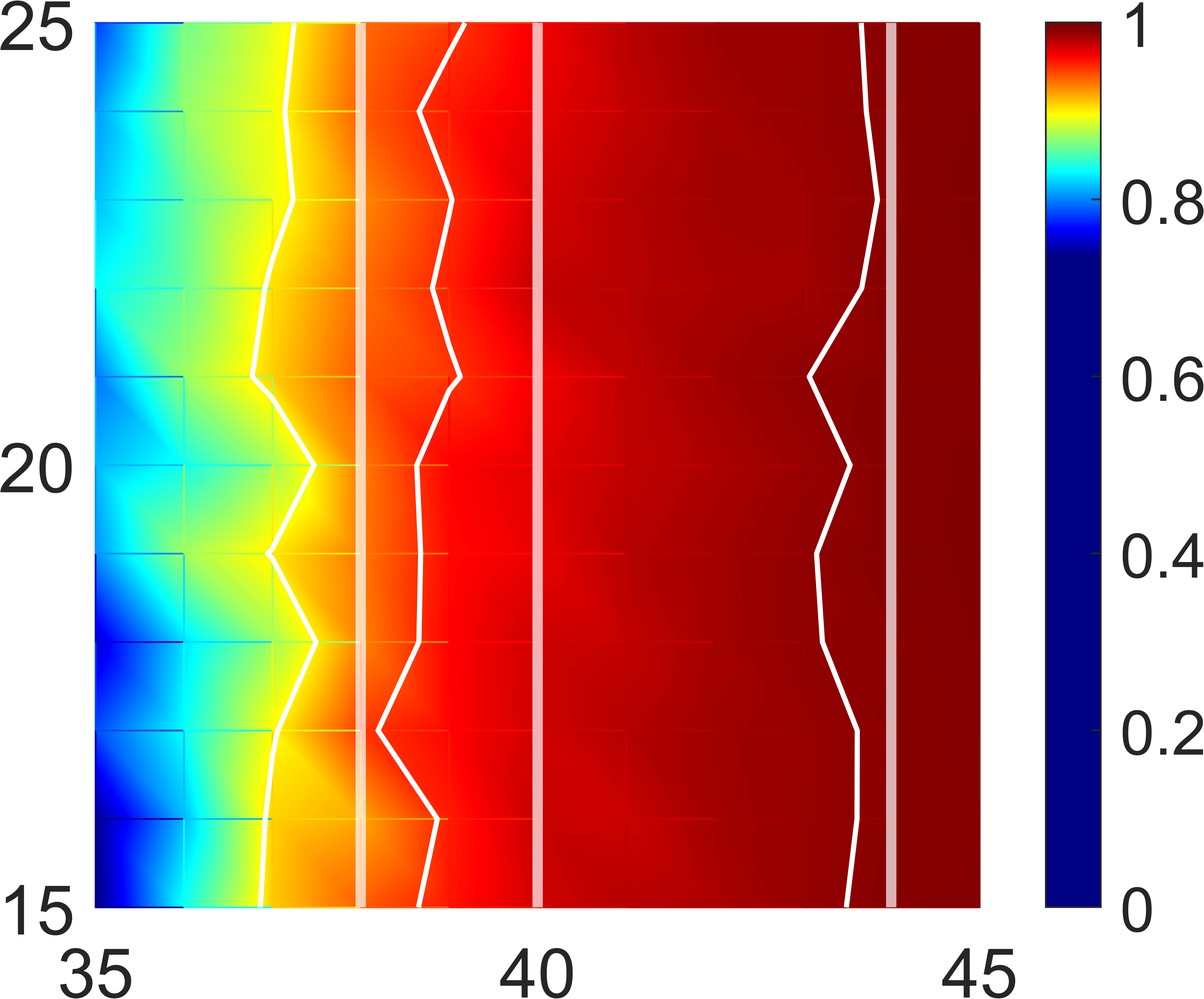} & 
\vspace{.3cm} \\
 & \centering $\xi=10$ &\centering $\xi=20$ &\centering $\xi=40$
\end{tabular}
\caption{\label{fig:chi_low_noise} Values of the fraction $\chi$ of successfully herd targets in the low noise case $D=10^{-2}$ obtained for different values of $M$ and $N$, each panel corresponding to different values of $R$ and $\xi$. Results are averaged over a number of simulations in the $[10:60]$ range, depending on the availability of computational resources. The increments of $N$ and $\sqrt{M}$ have values $\Delta N=1$, $\Delta \sqrt{M}=1$. The level curves for $\chi^*\in\{0.90,\,0.95,\,0.99\}$ are depicted by the white curves. The critical threshold $M>\mlow(\chi^*)$ are represented by the white vertical lines.}
\end{figure*}

In particular, for the   $D=10^{-2}$ case we found that the condition needed for the herders to balance the diffusion of the targets is met with a small number of herders, $N^*(M)$. In this scenario, the primary criterion  for herdability is that the number of targets, $M$, needs to exceed the threshold $\mlow$, as clearly depicted in Fig. \ref{fig:different_noise_50_10}a.
A preliminary scaling analysis for the observed $\mlow$, less extensive than the one discussed for the $D=1$ case, is reported in Fig. \ref{fig:scaling_supp_low_noise}, where the values of $\mlow$ are obtained empirically by direct inspection of \ref{fig:chi_low_noise} (all the values of $R$, $\xi$, and $\mlow(\chi^*)$ are reported in Table \ref{tab:table2}).

\section{Random Geometric Graphs and the estimate of $\widehat{\mlow}$}
\label{mysection:percolation}

Random geometric graphs (RGGs) have been extensively studied in the literature, e.g. \cite{BARTHELEMY20111,dall2002random}. An RGG is constructed by distributing nodes within a specified domain—often a two-dimensional area—and connecting pairs of nodes with an edge based on a predetermined geometric criterion. The most fundamental of these criteria is the proximity rule, wherein an edge is formed between any two nodes if their Euclidean distance falls below a specified threshold, denoted by $r$.

In the context of our study, we introduce the concept of the herdability graph, $\mathcal{G}$, as a unique instantiation of an RGG. Here, the nodes represent targets positioned within a planar domain, and edges are established between targets $\textbf{T}_a$ and $\textbf{T}_b$ if the distance between them, $|\textbf{T}_a-\textbf{T}_b|$, does not exceed $\xi$. This definition aligns $\mathcal{G}$ with the characteristics of a random geometric graph, adopting $\xi$ as the equivalent threshold parameter, $r$.

Given a finite domain $\mathcal{D}\in\R^2$ where $M$ nodes are randomly and uniformly distributed, we can define the total excluded volume (or connectivity) $\alpha$ \cite{dall2002random} as $\alpha=M (\pi r^2)/|\mathcal{D}|$; it is then known from the literature that by increasing $\alpha$, the corresponding random geometric graph undergoes a percolation at a critical value, say $\alpha_c$ \cite{dall2002random}. 

To demonstrate that the percolation density of $\mathcal{G}$ serves as an effective estimator for the empirically observed $\mlow$, we consider the worst-case scenario at the initial time $t=0$. Here, $M$ targets are randomly and uniformly distributed within a circle $\oo$ of radius $R$. We can then define the total excluded volume as $M (\xi/R)^2$ and look for the percolation threshold of the herdability graph.

\begin{figure}[h]
    \includegraphics[width=.5\textwidth]{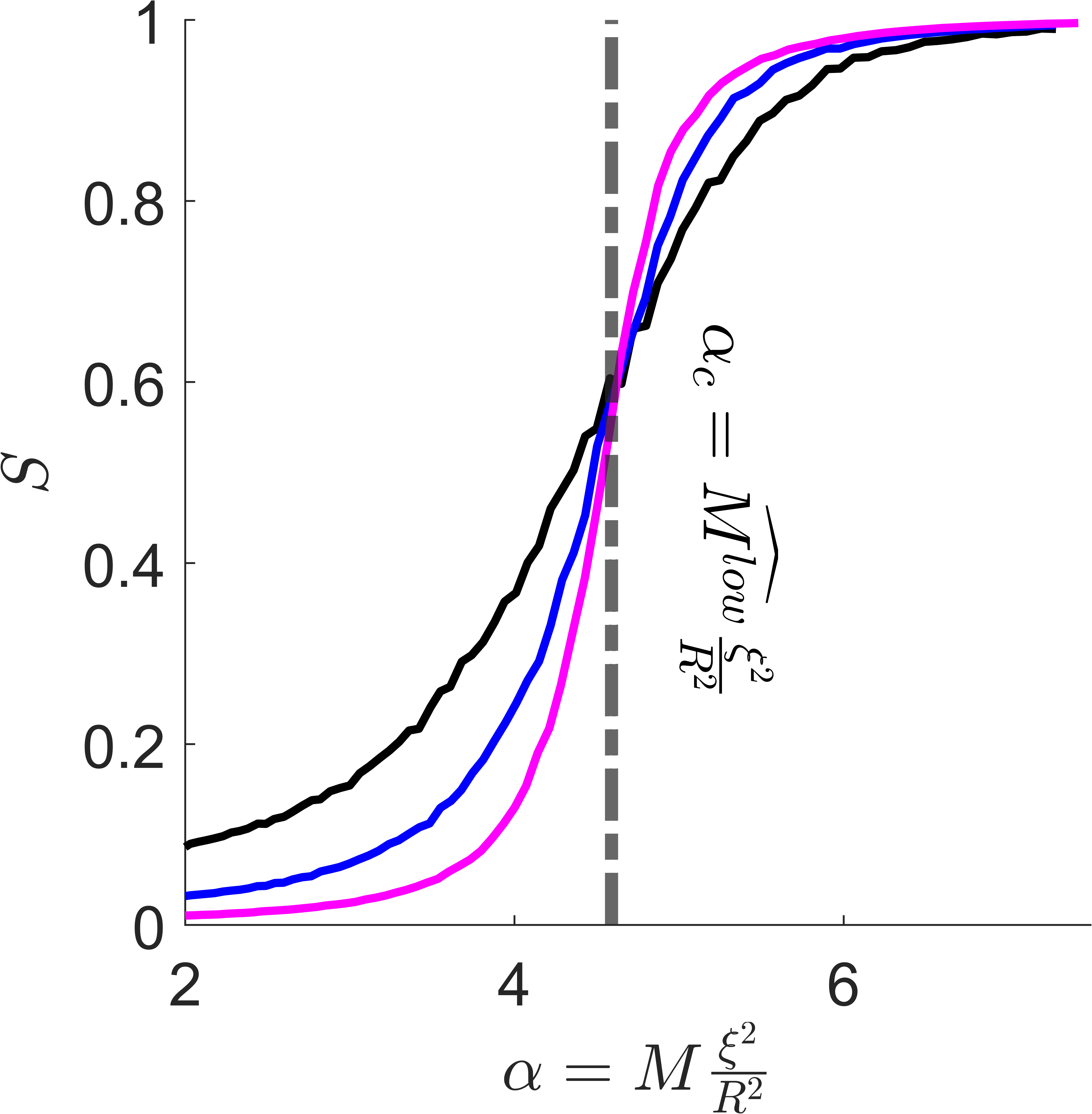}
\caption{\label{fig:perc} Percolation analysis of $\mathcal{G}$. Here we plot the fraction of nodes in the largest connected component of $\mathcal{G}$, $S$, as a function of the connectivity $\alpha=M(\xi/R)^2$. We consider $\xi=10$ and different values of $R$ (black, blue, and magenta curve for $R=100$, $R=200$, and $R=400$ respectively). The percolation threshold $\alpha_c$ (black dashed line) is chosen as the point where the different curves intersect. Each point is the average over $500$ simulations.}
\end{figure}

In Fig. \ref{fig:perc} we report the fraction of nodes in the largest component $S$ of the herdability graph; the results are obtained via numerical simulations as a function of the connectivity $\alpha = M (\xi/R)^2$  for different values of $R$. We identify empirically the percolation threshold $\alpha_c \approx 4.60$ as the value at which curves at different $R$ values cross \cite{cho2010finite}. Finally, by defining $\widehat{\mlow}(\xi,R)$ the critical number of nodes at which the connectivity $\alpha=\alpha_c$, we get
\begin{equation}
\widehat{\mlow}(\xi,R)=\alpha_c \left(\dfrac{R}{\xi}\right )^2=4.60 \left(\dfrac{R}{\xi}\right )^2
\label{eqn:percolation}
\end{equation}
an estimate accurate enough for our purpose and in accordance with \cite{dall2002random}.

\section{Effects of Periodic Boundary Conditions}
\begin{figure}[h]
    \centering
    \begin{overpic}[width=.55\textwidth]{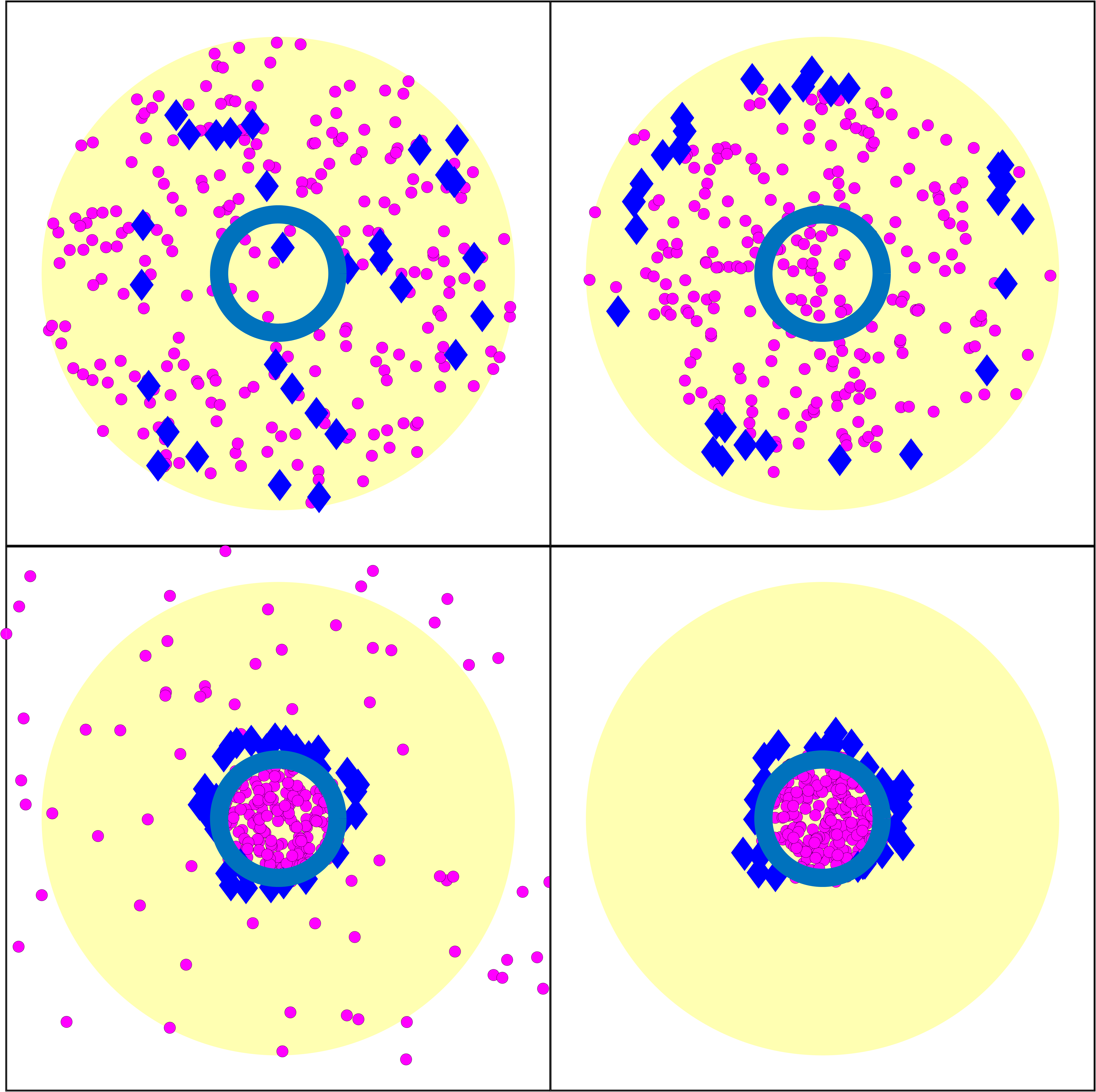}\vspace{.5cm}
    \put(2,96){(a)}
    \put(94,96){(b)}
    \put(2,2){(c)}
    \put(94,2){(d)}
    \end{overpic}
    \\ \vspace{0.5cm}
    \begin{overpic}[width=.6\textwidth]{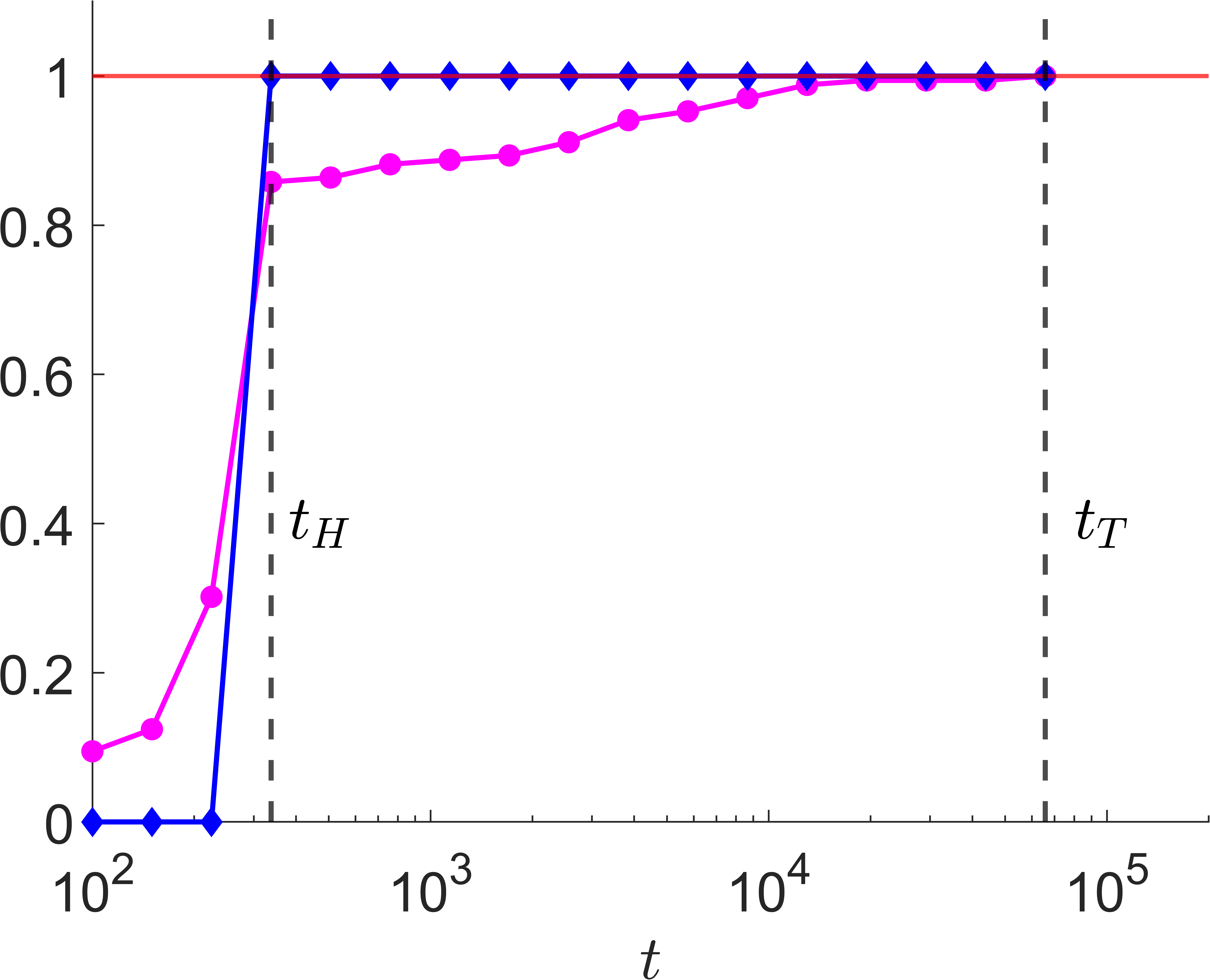}
    \put(2,79){(e)}

    \end{overpic}
    \caption{(a-d) Representative snapshots of the system when the Shepherding dynamics is simulated in a periodic box of finite size $L>R$ in the $M<\mlow$ regime. (a) At the initial time $t=0$  the agents are uniformly distributed in $\oo$. (b) At an intermediate time during shepherding control, the herders surround part of the targets, while a fraction of the targets is lost (diffuse beyond the herders) since  in the $M<\mlow$ regime the local information of the herders is not enough to timely observe, and hence chase, all the targets as discussed in the main.  (c) Eventually the herders arrive to $\og$ by containing therein the fraction of the targets they were able to observe; the fraction of the targets which were not observed by the herders is now diffusing in the domain. (d) Since the domain is a periodic box of finite size, the targets will eventually diffuse towards $\og$ and will be then captured by the herders.
    Hence, with periodic boundary conditions, even in the $M<\mlow$ case, the herders will eventually be able after a finite time to encircle all the targets. 
    This finding distinguishes our study from existing literature on phase separations, where phenomena resembling shepherding are observed. Notably, in our model, the dilute phase of passive-targets always disappears. (e) Fraction of targets $\chi$ and of herders $\psi$ in the goal region $\og$ as a function of time (magenta and blue lines respectively); the red line highlights the value $1$. At a time $t_H$ (left vertical dashed line) all the herders arrive on $\og$; the remaining fraction of the targets, that at $t=t_H$ were not observed and captured by the herders, will eventually diffuse towards $\og$ after a time $t_T>t_H$ (right vertical dashed line).}
    \label{fig:PBC}
\end{figure}

This section demonstrates the impact of periodic boundary conditions (PBCs) in a box, a common choice in studies of active-passive mixtures where phase separations occur \cite{dolai2018phase,stenhammar2015activity}. As depicted in Fig.\ref{fig:PBC}, under PBCs, herders can eventually encircle all targets, even in the $M<\mlow$ regime. This is because targets that might otherwise escape in non-periodic conditions will ultimately diffuse towards the goal region in a finite time and be intercepted by the herders. This finding distinguishes our study from existing literature on phase separations, where phenomena resembling shepherding are observed. Notably, in our model, the dilute phase of passive-targets always disappears, without necessitating specific initial conditions as in previous studies (e.g., \cite{stenhammar2015activity}).

\newpage
\section{Supplementary tables}

\begin{table}[ht]
\centering
\begin{tabular}{|l|c|c|c|r|}
\hline
\rule{0pt}{2.5ex}$R$ & $\xi$ & $\sqrt{\mlow}(\chi^*=0.90)$ & $\sqrt{\mlow}(\chi^*=0.95)$ & $\sqrt{\mlow}(\chi^*=0.99)$\\
\hline
50 & 10 & 8 & 9 & 11 \\
\hline
100 & 20 & 8 & 9 & 12 \\
\hline
100 & 10 & 16 & 18 & 22 \\
\hline
200 & 40 & 9 & 10 & 12 \\
\hline
200 & 20 & 17 & 19 & 25 \\
\hline
200 & 10 & 31 & 35 & 45 \\
\hline
400 & 80 & 9 & 10 & 12 \\
\hline
400 & 40 & 18 & 19 & 23 \\
\hline
400 & 20 & 34 & 36 & 44 \\
\hline
400 & 10 & 61 & 65 & 83 \\
\hline
\end{tabular}
\caption{Values of $\sqrt{\mlow}(\chi^*)$, with $\chi^*\in\{0.90,\,0.95,\,0.99\}$, for different values of $R$ and $\xi$. These data are used to realize Fig. \ref{fig:scaling_supp}}
\label{tab:table1}
\end{table}

\begin{table}[!h]
    \centering
    \begin{tabular}{|l|c|c|c|r|}
    \hline
\rule{0pt}{2.5ex}    $R$ & $\xi$ & $\sqrt{\mlow}(\chi^*=0.90)$& $\sqrt{\mlow}(\chi^*=0.95)$& $\sqrt{\mlow}(\chi^*=0.99)$\\
    \hline
       50   & 10 & 9  & 10 & 12 \\
           \hline
       100  & 20 & 9 & 10 & 13 \\
           \hline
       100  & 10 & 18 & 20 & 23 \\
           \hline
       200  & 40 & 10 & 11 & 13 \\
           \hline
       200  & 20 & 18 & 20 & 23 \\
           \hline
       200  & 10 & n/a  & n/a  & 43 \\
           \hline
       400  & 20 & 38 &40 & 44 \\
       \hline
    \end{tabular}
    \caption{Values of $\sqrt{\mlow}(\chi^*)$, with $\chi^*\in\{0.90,\,0.95,\,0.99\}$, for different values of $R$ and $\xi$ for the  low noise case $D=10^{-2}$. These data are used to realize Fig. \ref{fig:scaling_supp_low_noise}}
    \label{tab:table2}
\end{table} 

\newpage

\section{Supplementary videos}

\begin{video}[ht]
\centering
\includegraphics[width=0.5\textwidth]{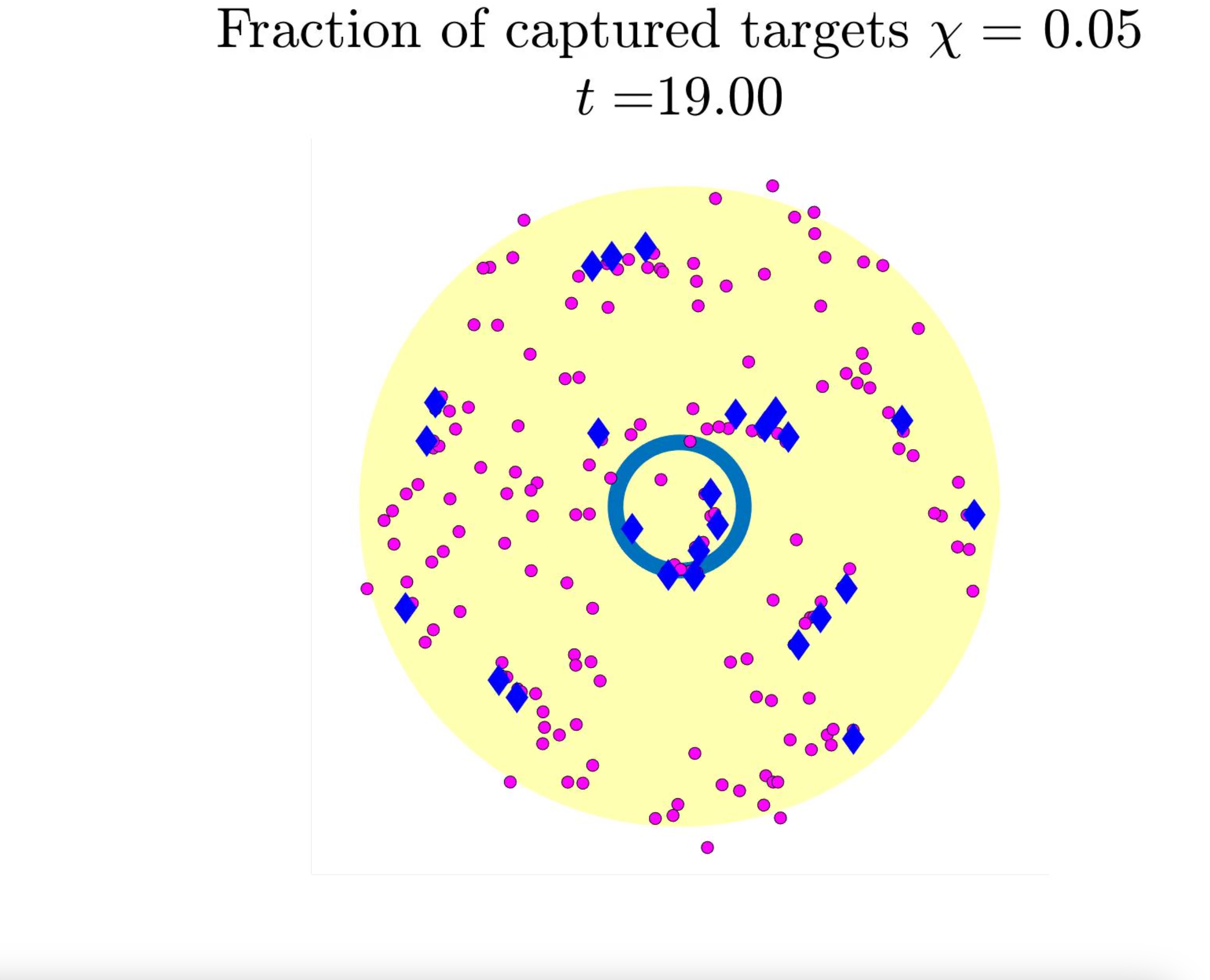} 
\caption{Supplementary Video 1: Representative evolution of the herders and targets dynamics when $M<\mlow$ and the conditions for successful herding are not met. For the full video, see \href{https://www.dropbox.com/scl/fi/65en0rbkv79l0shruoz3g/FAIL-1.mp4?rlkey=82r8e3eio4gecwr7h2c51raht&dl=0}{Video Link}.}
\label{fig:video1}
\end{video}

\begin{video}[ht]
\centering
\includegraphics[width=0.5\textwidth]{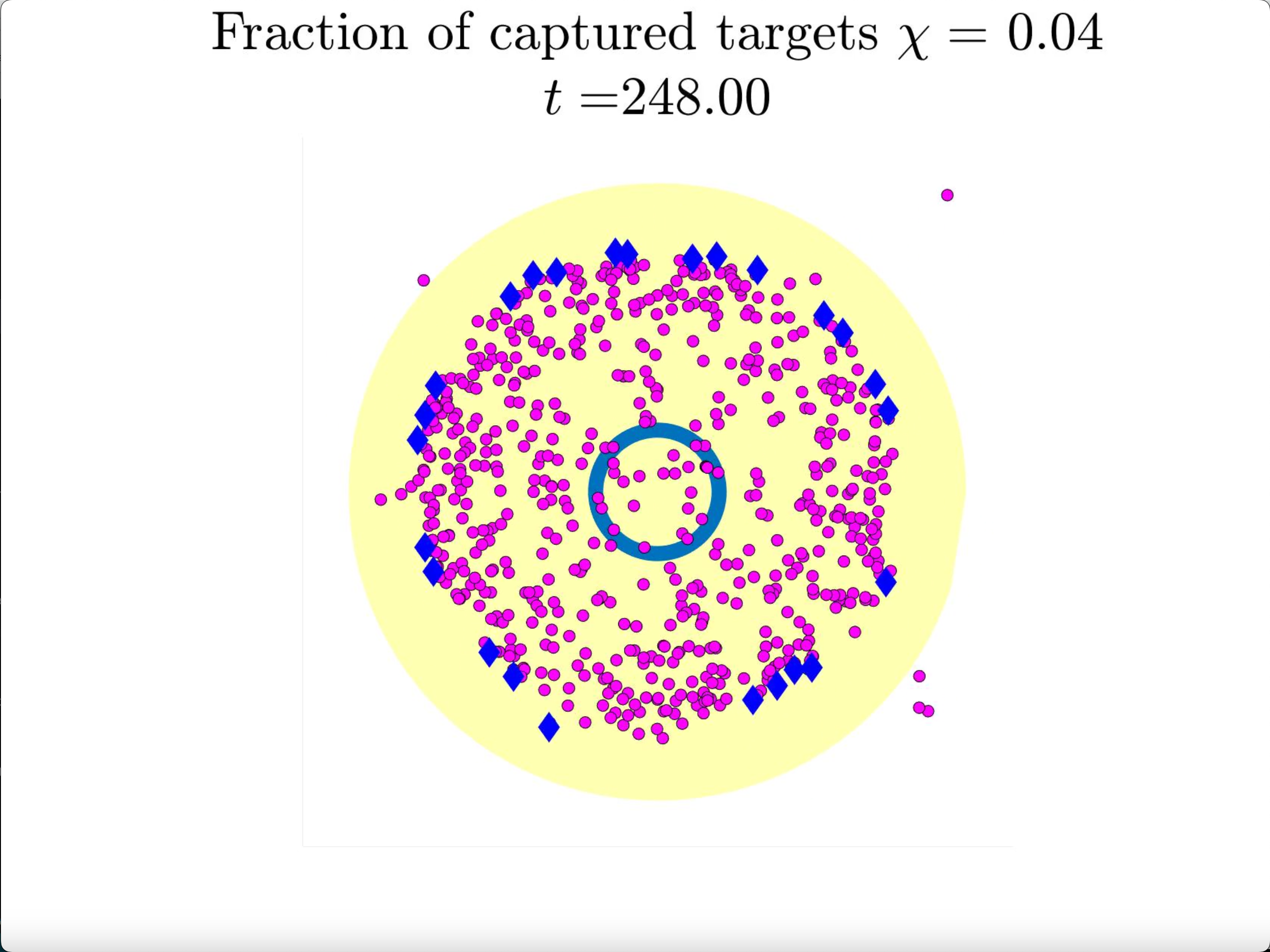} 
\caption{Supplementary Video 2: Representative evolution of the herders and targets dynamics when $M>\mlow$ and the conditions for successful herding are met. For the full video, see \href{https://www.dropbox.com/scl/fi/a2fn1gziex0ros7hyt35n/SUCC-1.mp4?rlkey=flnnkqidj2xc1t0uu0wfoc8ob&dl=0}{Video Link}.}
\label{fig:video1}
\end{video}

\newpage

\section{Supplementary figures}

\begin{figure*}[h]
    \begin{tabular}{lllll}
$R=50$ & \includegraphics[width=.2\linewidth,valign=m]{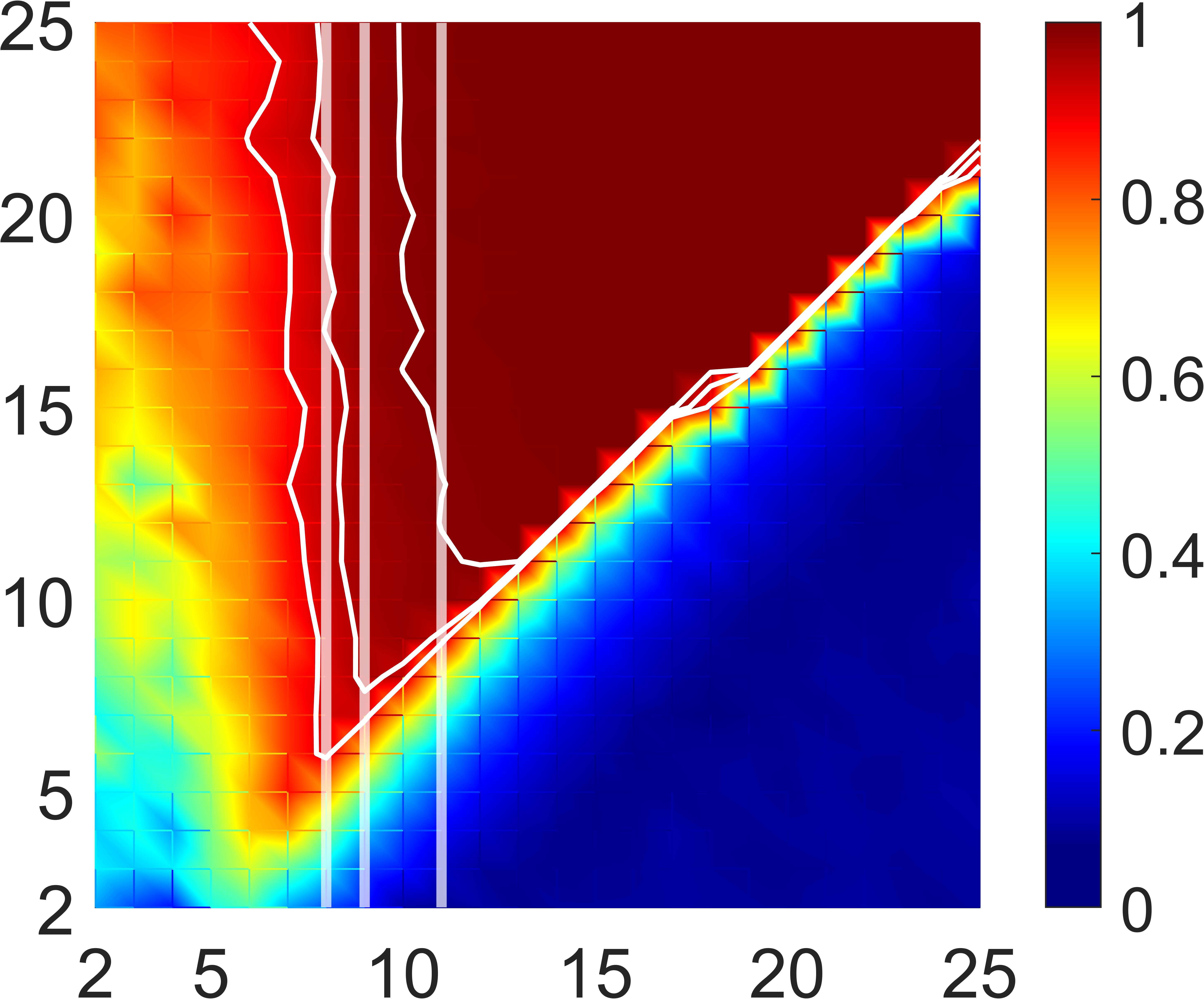} &  &  \\
$R=100$ & \includegraphics[width=.2\linewidth,valign=m]{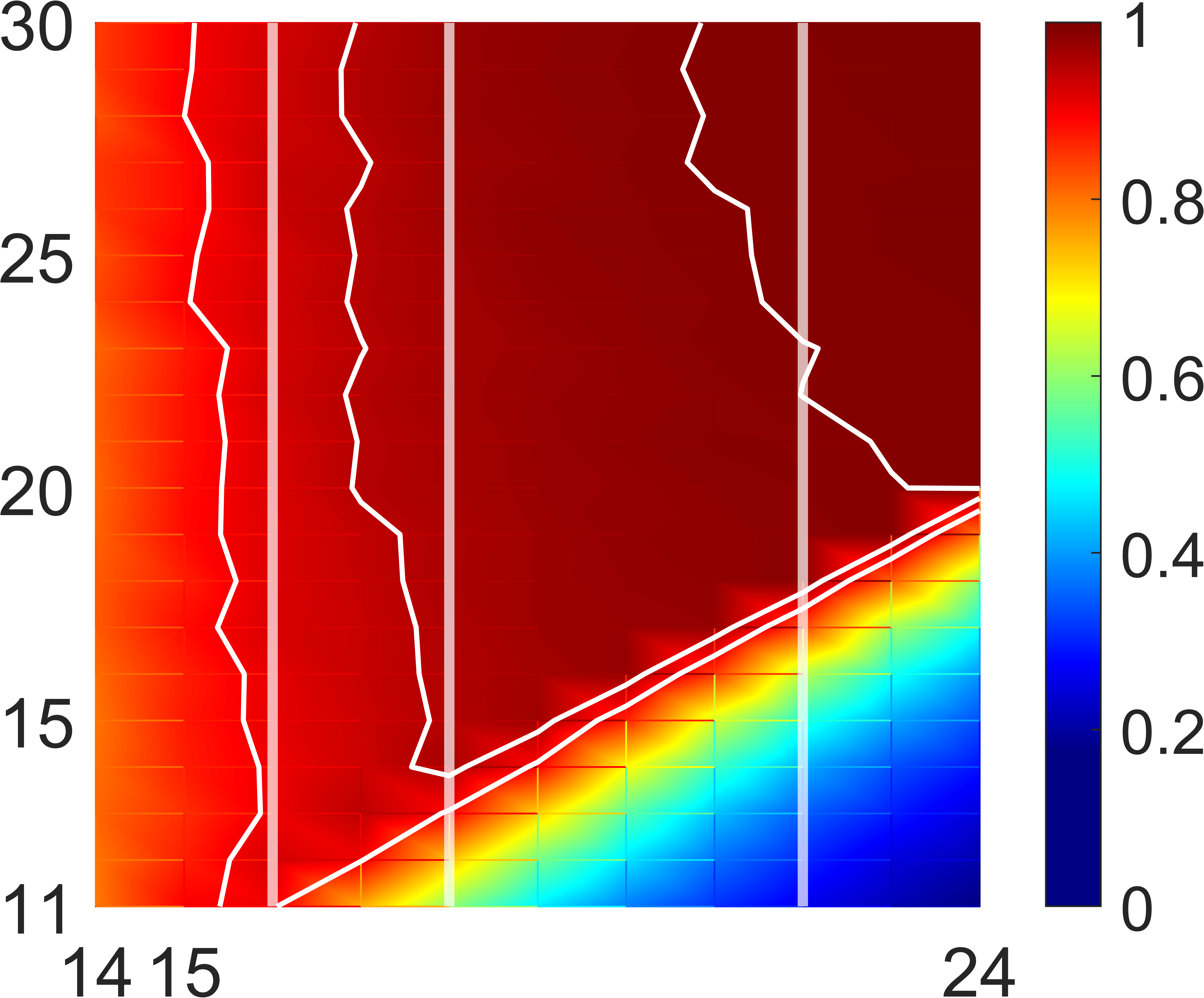} & \includegraphics[width=.2\linewidth,valign=m]{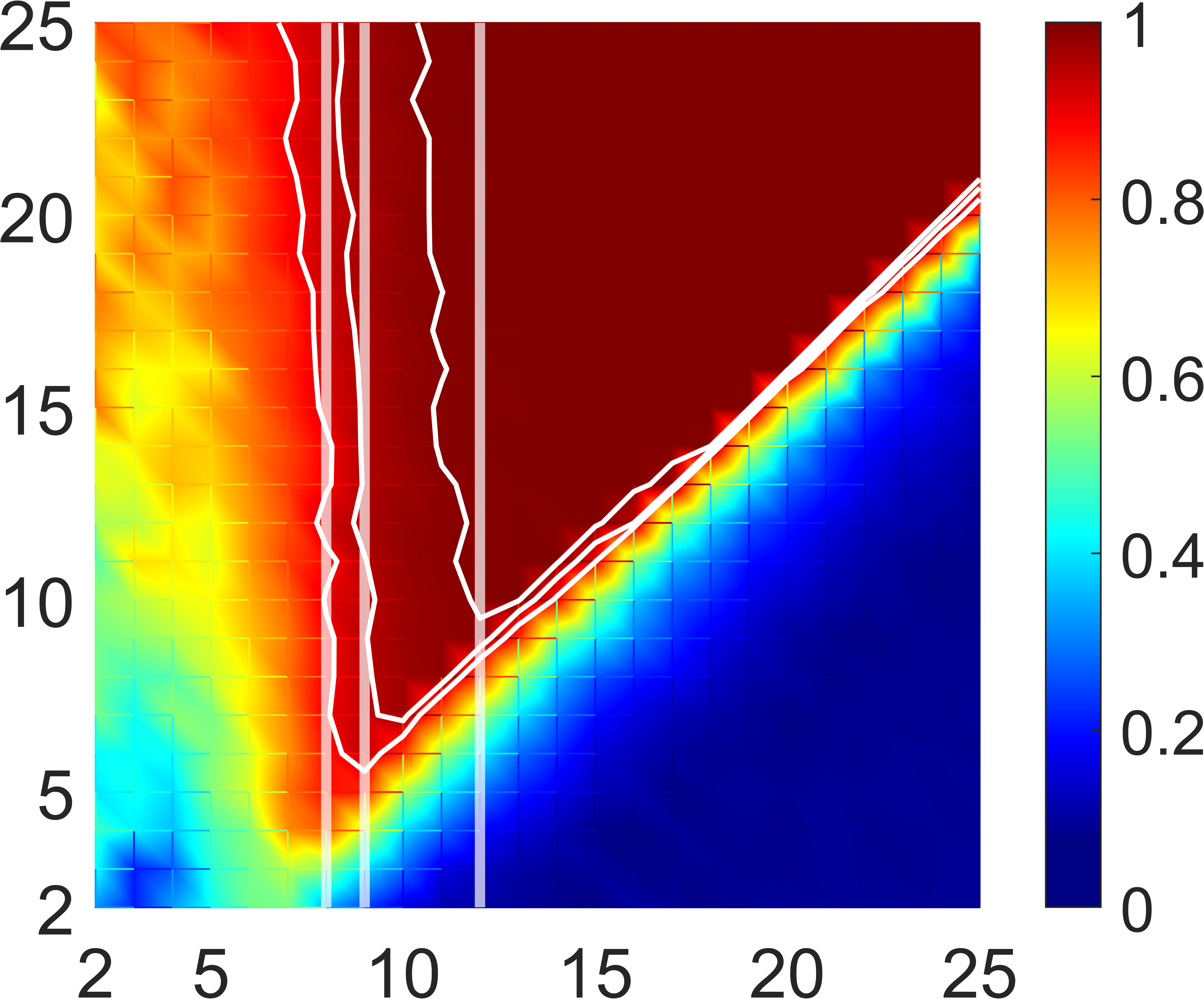} &  &  \\
$R=200$ & \includegraphics[width=.2\linewidth,valign=m]{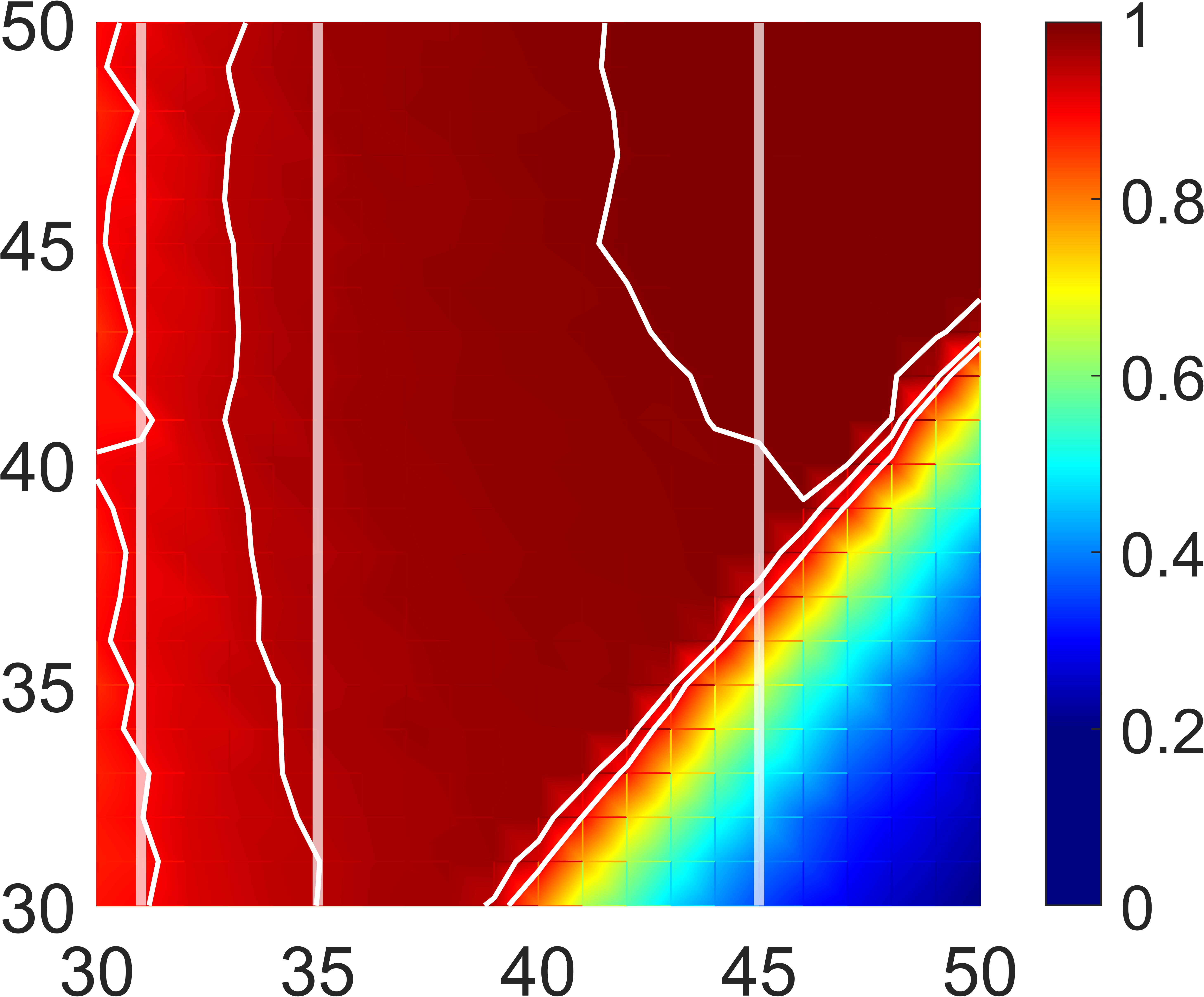} & \includegraphics[width=.2\linewidth,valign=m]{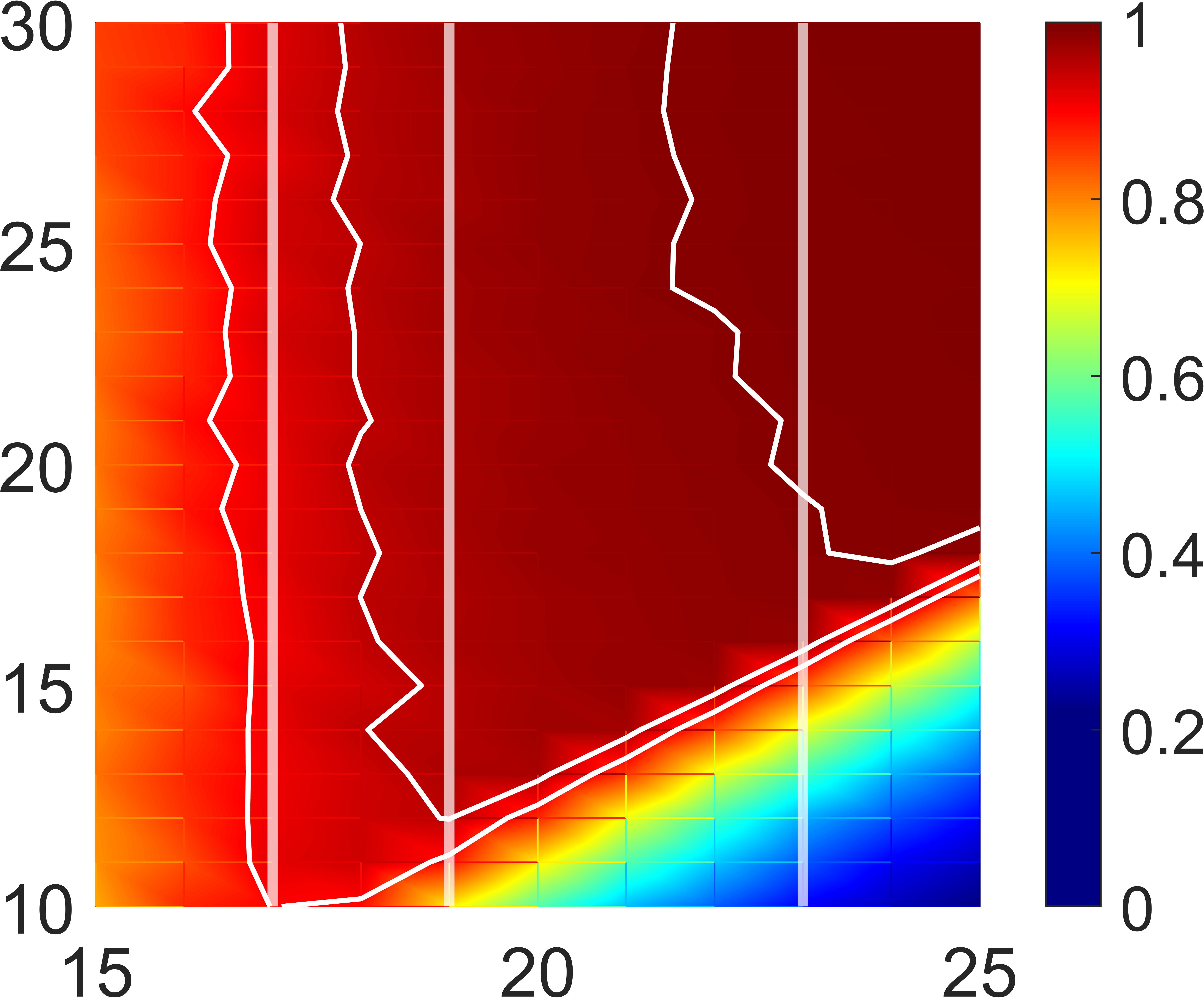} & \includegraphics[width=.2\linewidth,valign=m]{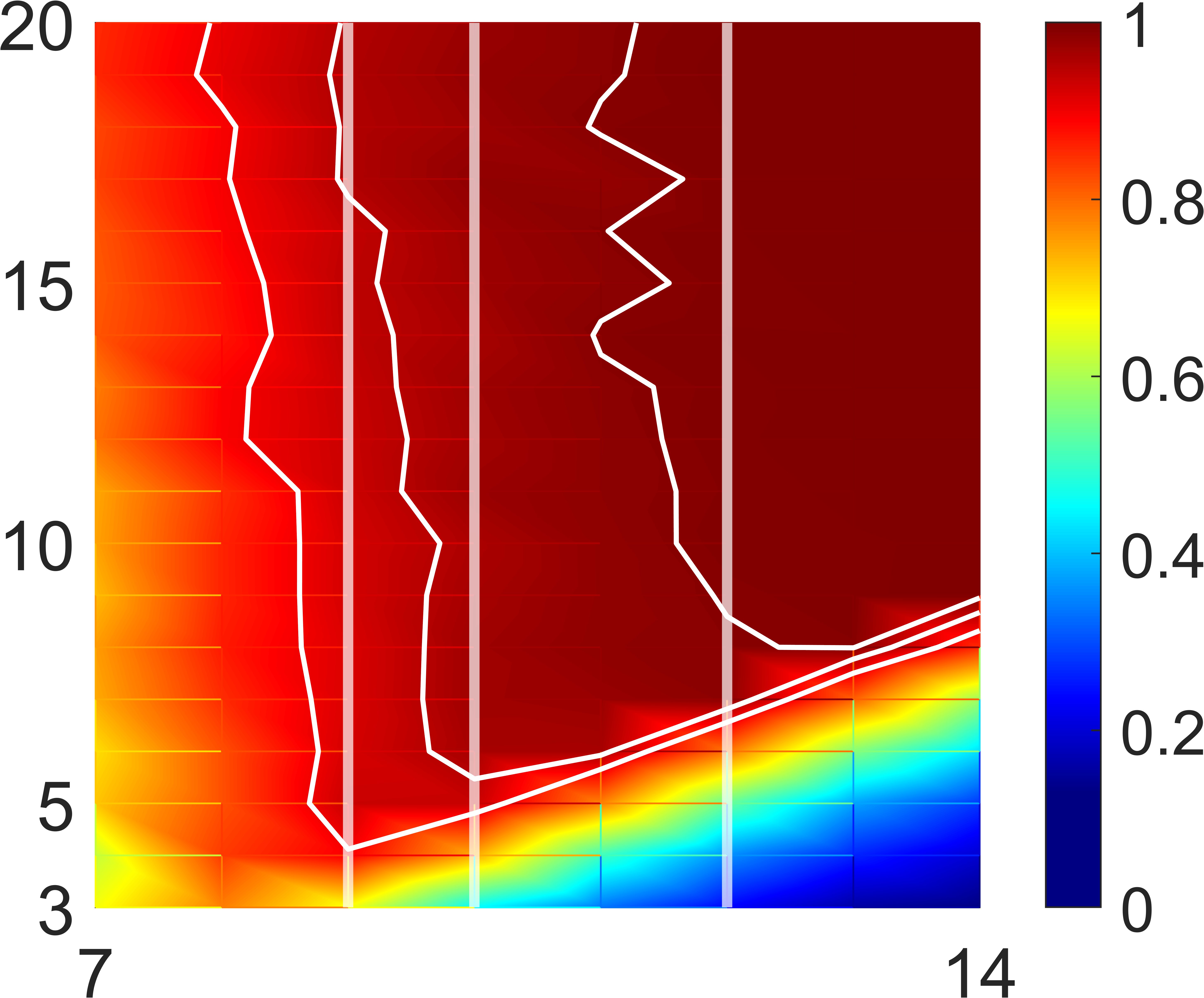}&
\\
$R=400$ & \includegraphics[width=.2\linewidth,valign=m]{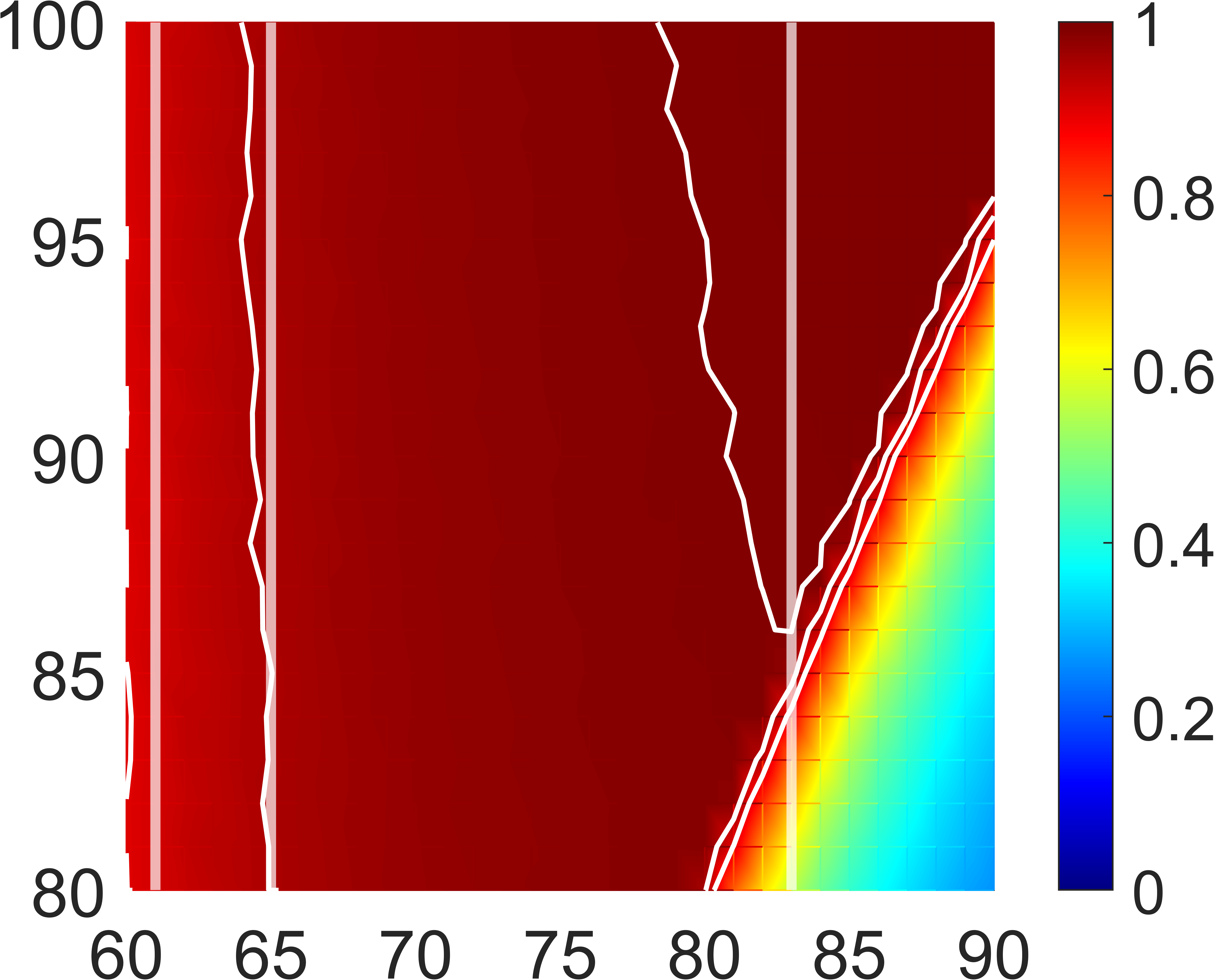} & \includegraphics[width=.2\linewidth,valign=m]{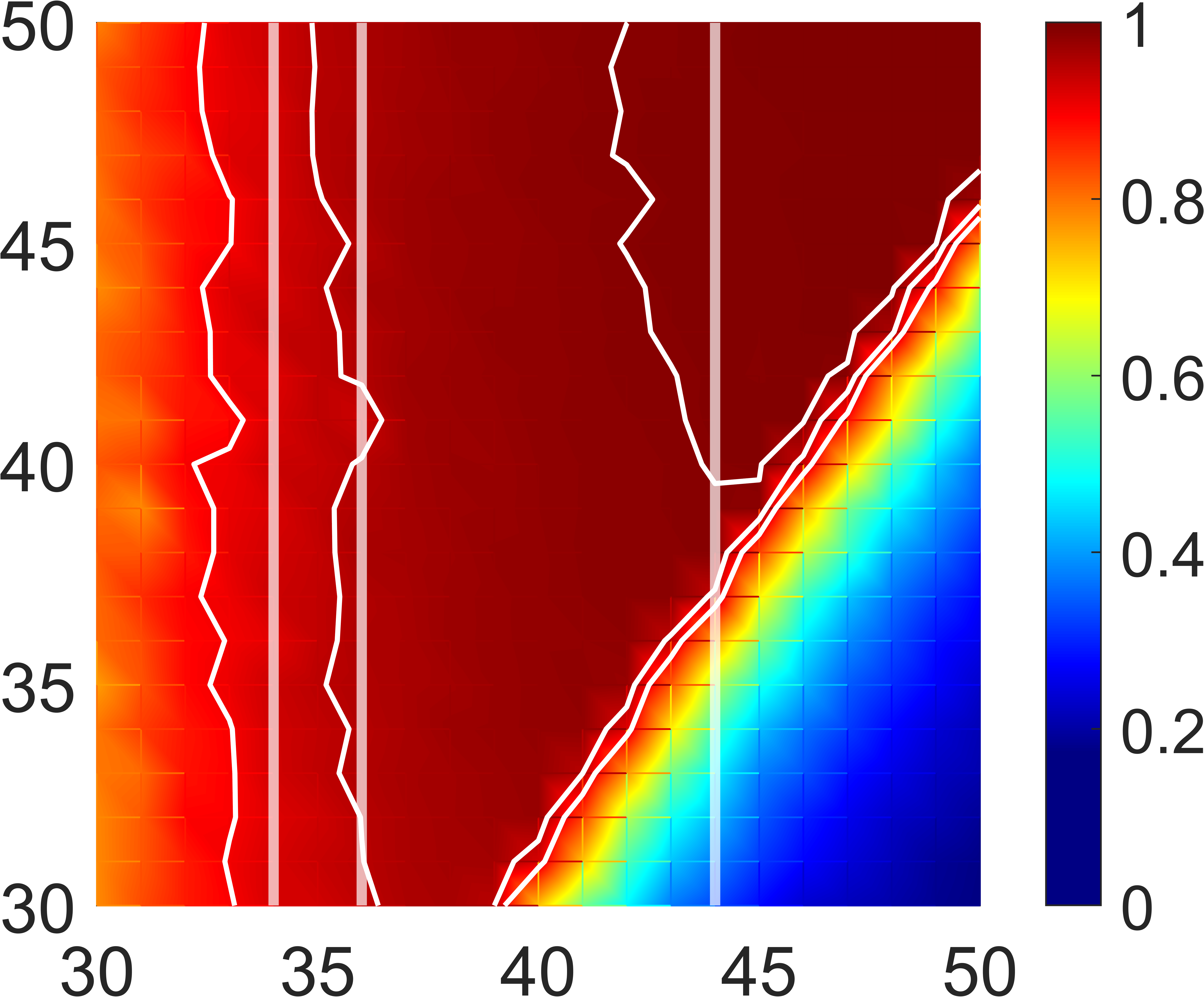} & \includegraphics[width=.2\linewidth,valign=m]{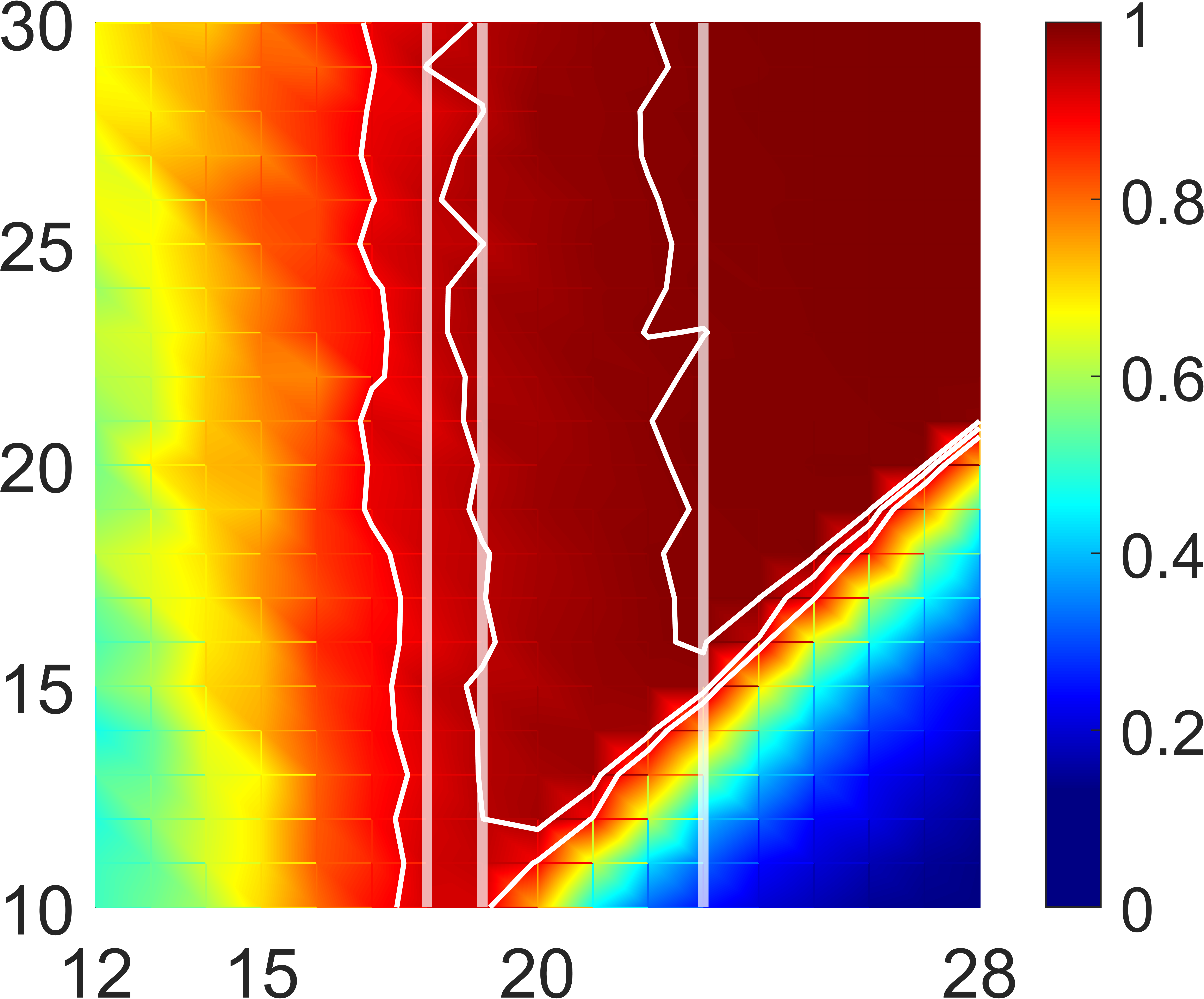}&
\includegraphics[width=.2\linewidth,valign=m]{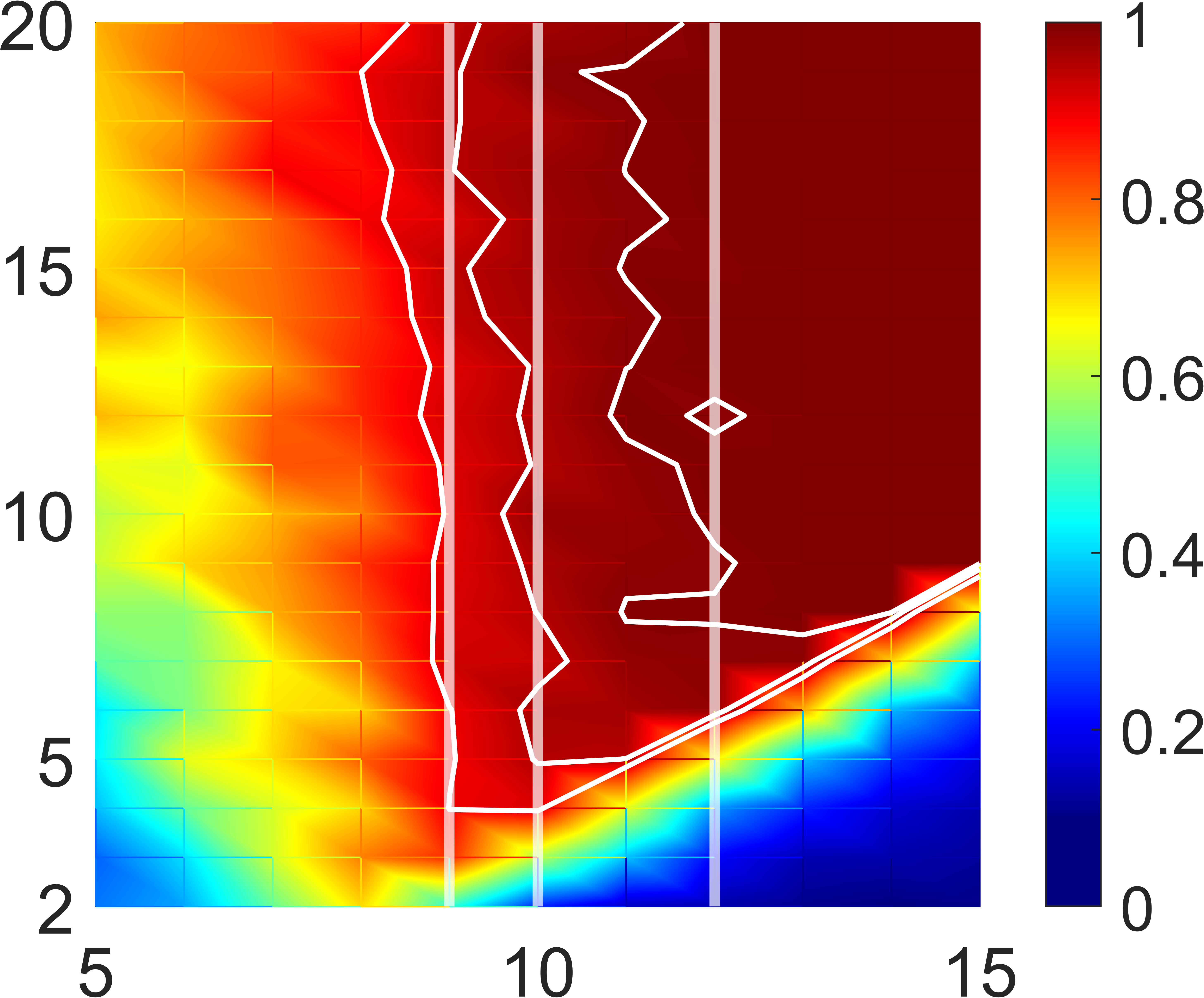}
\vspace{.3cm}\\
 & \centering $\xi=10$ &\centering $\xi=20$ &\centering $\xi=40$ &\centering $\xi=80$
\end{tabular}
\caption{\label{fig:chi} Values of the fraction $\chi$ of successfully herd targets obtained for different values of $M$ and $N$, each panel corresponding to different values of $R$ and $\xi$. Results are averaged over a number of simulations in the $[10:60]$ range, depending on the availability of computational resources. The increments of $N$ and $\sqrt{M}$ have values $\Delta N=1$, $\Delta \sqrt{M}=1$. The level curves for $\chi^*=0.90$, $\chi^*=0.95$ and $\chi^*=0.99$, respectively, are depicted by the white curves. The corresponding empirically computed critical thresholds $\mlow(\chi^*)$ are represented by the white vertical lines.}
\end{figure*} 

\begin{figure*}[h]
\centering
        \begin{overpic}[width=.45\textwidth]{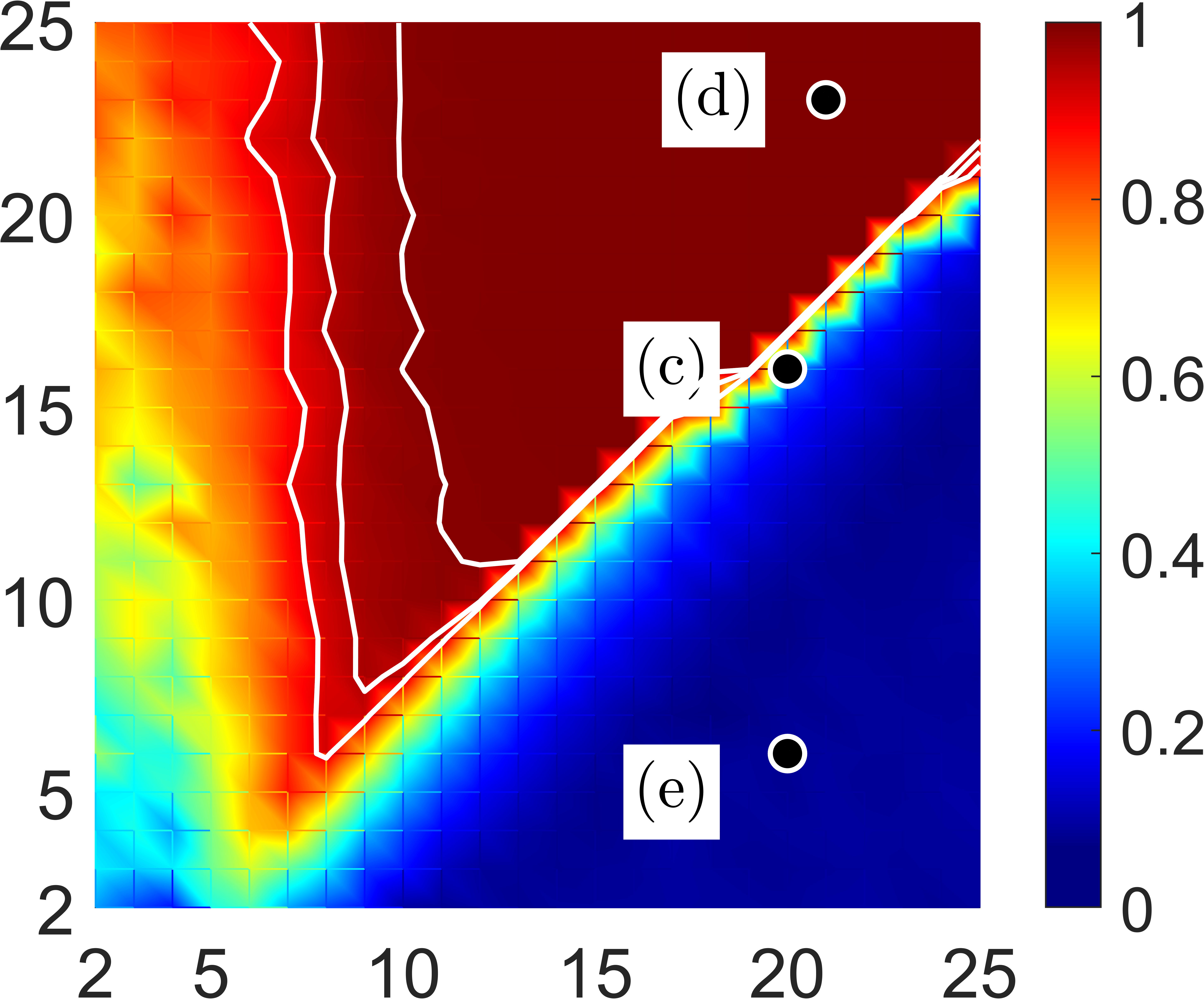}
    \put(2,88){(a)}
    \put(40,-6){$\sqrt{M}$}
    \put(-6,40){$N$}
    \put(40,85){$\chi$}
    \end{overpic}
    \begin{overpic}[width=.45\textwidth]{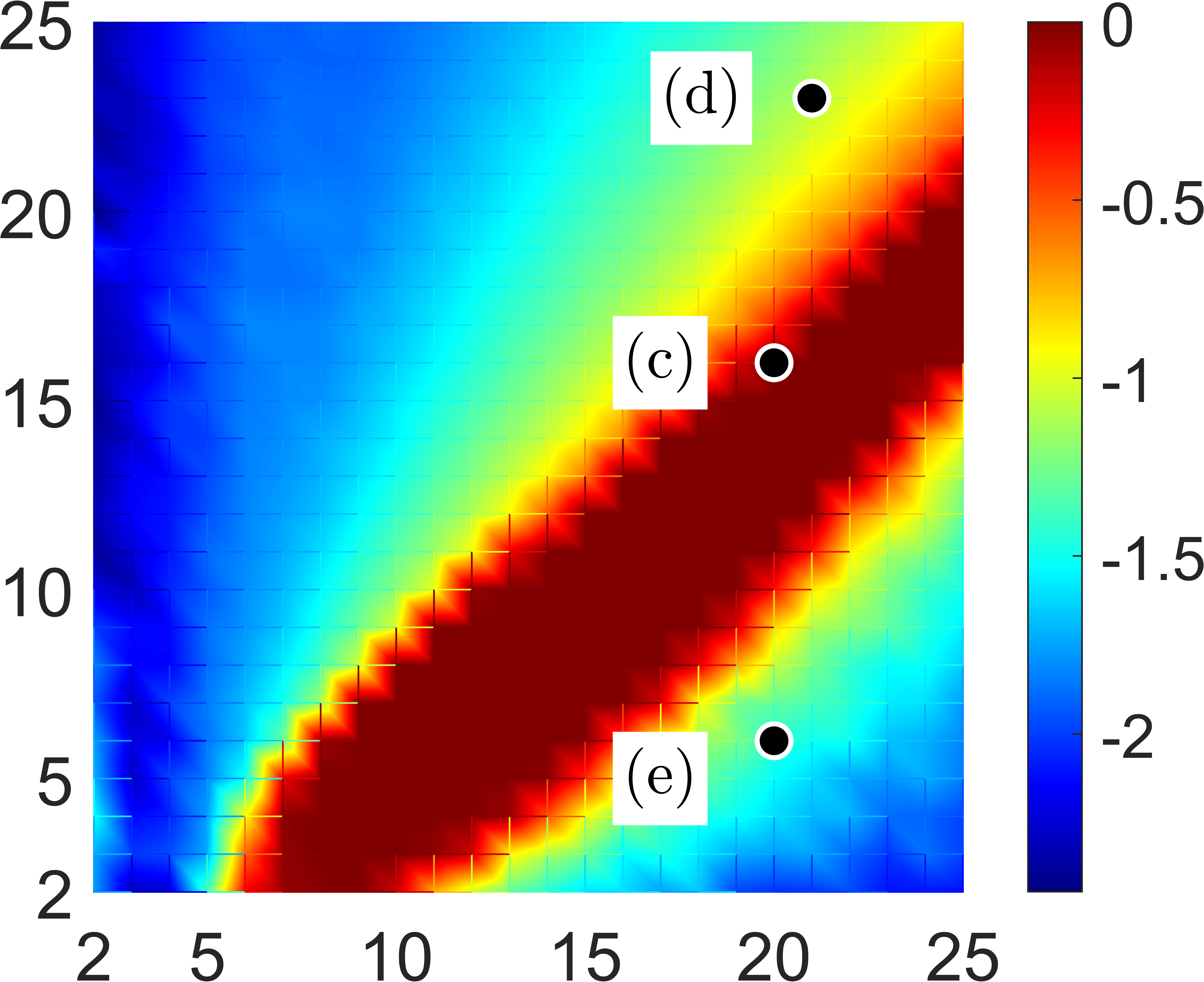}
    \put(2,88){(b)}
    \put(42,-6){$\sqrt{M}$}
    \put(-6,40){$N$}
        \put(33,85){$\log (t^{end}/t^{max})$}
    \end{overpic}  
    \\ \vspace{1.5cm}
    \begin{overpic}[width=.3\textwidth]{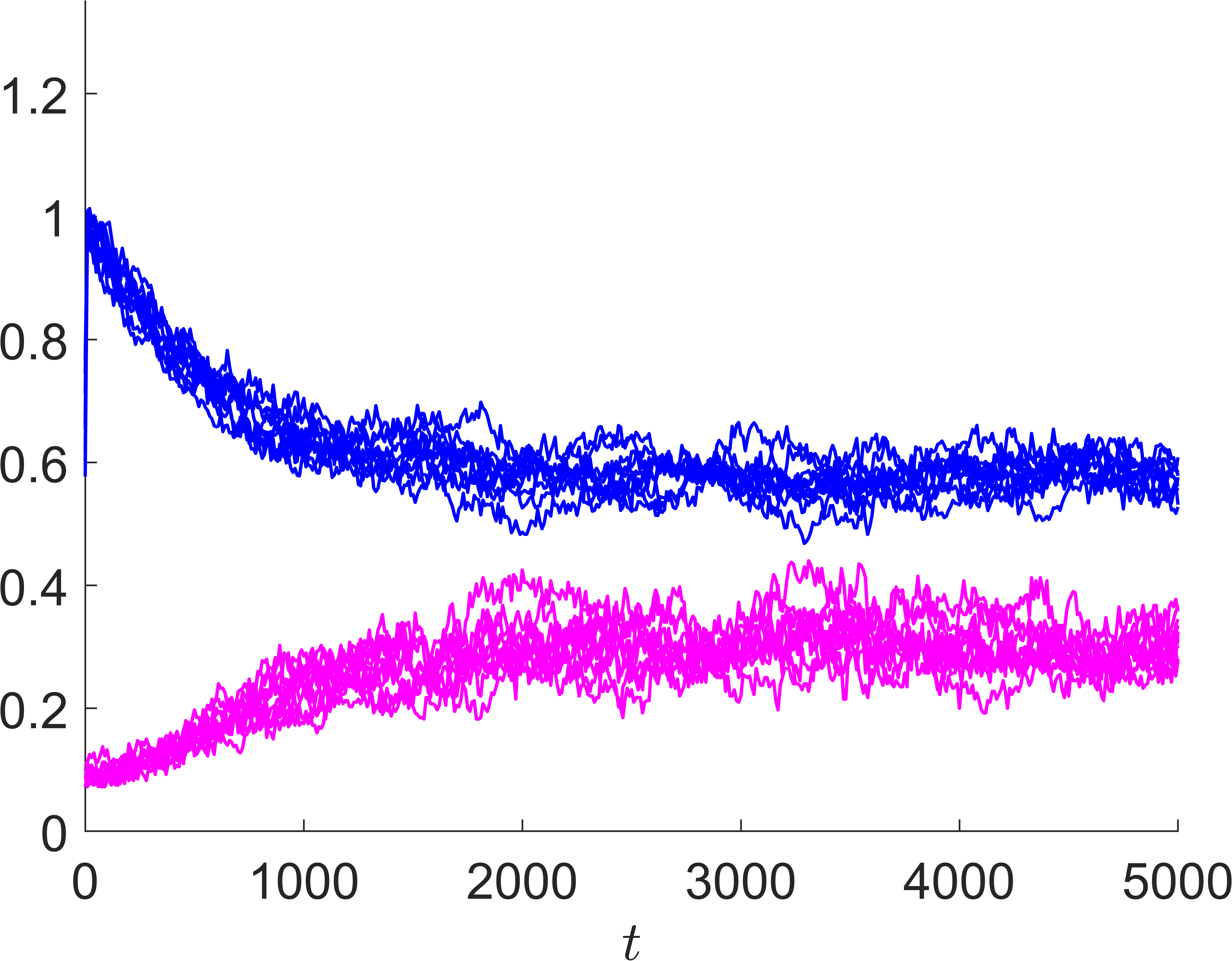}
    \put(2,85){(c)}
    \end{overpic}
    \begin{overpic}[width=.3\textwidth]{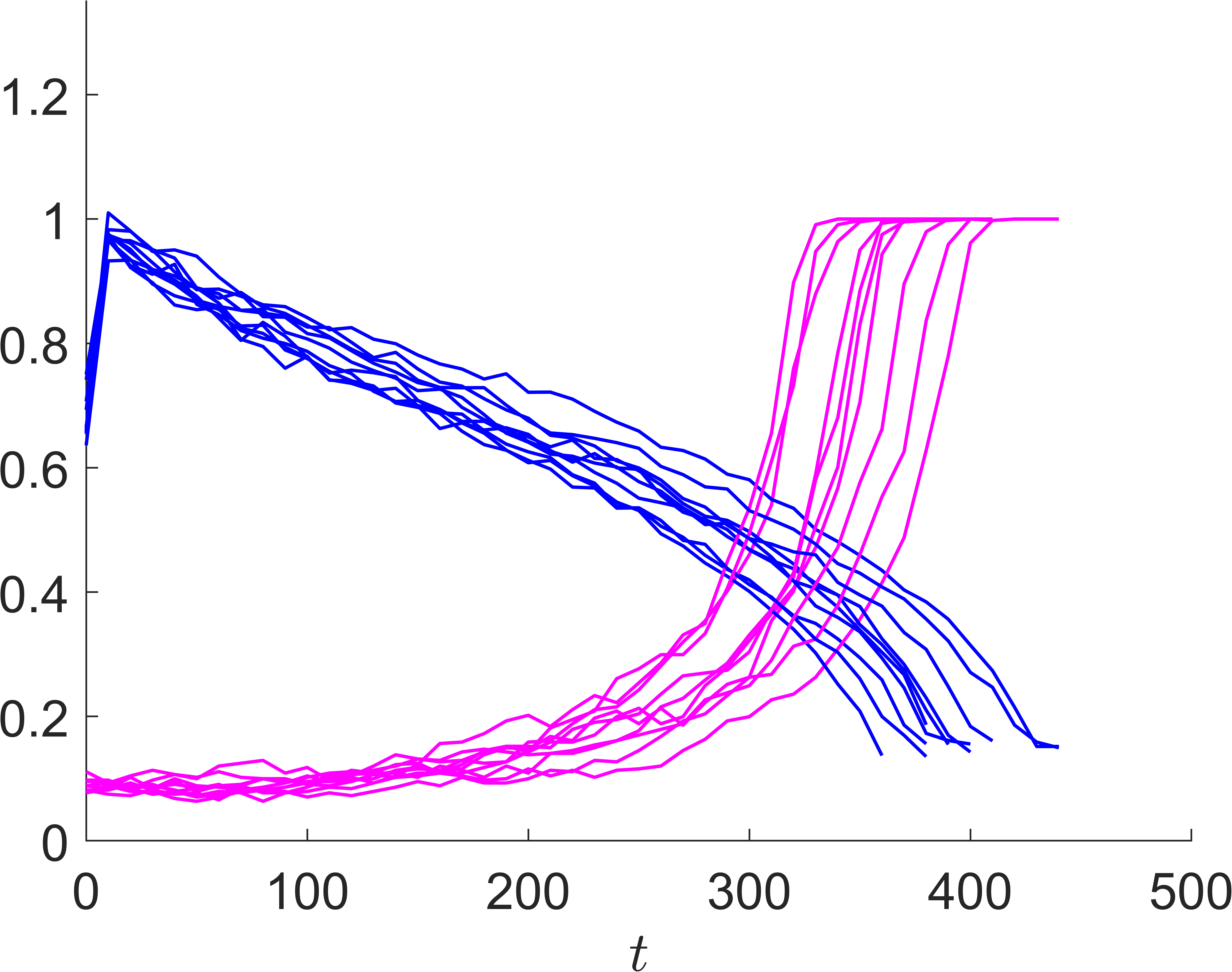}
    \put(2,85){(d)}
    \end{overpic}
    \begin{overpic}[width=.3\textwidth]{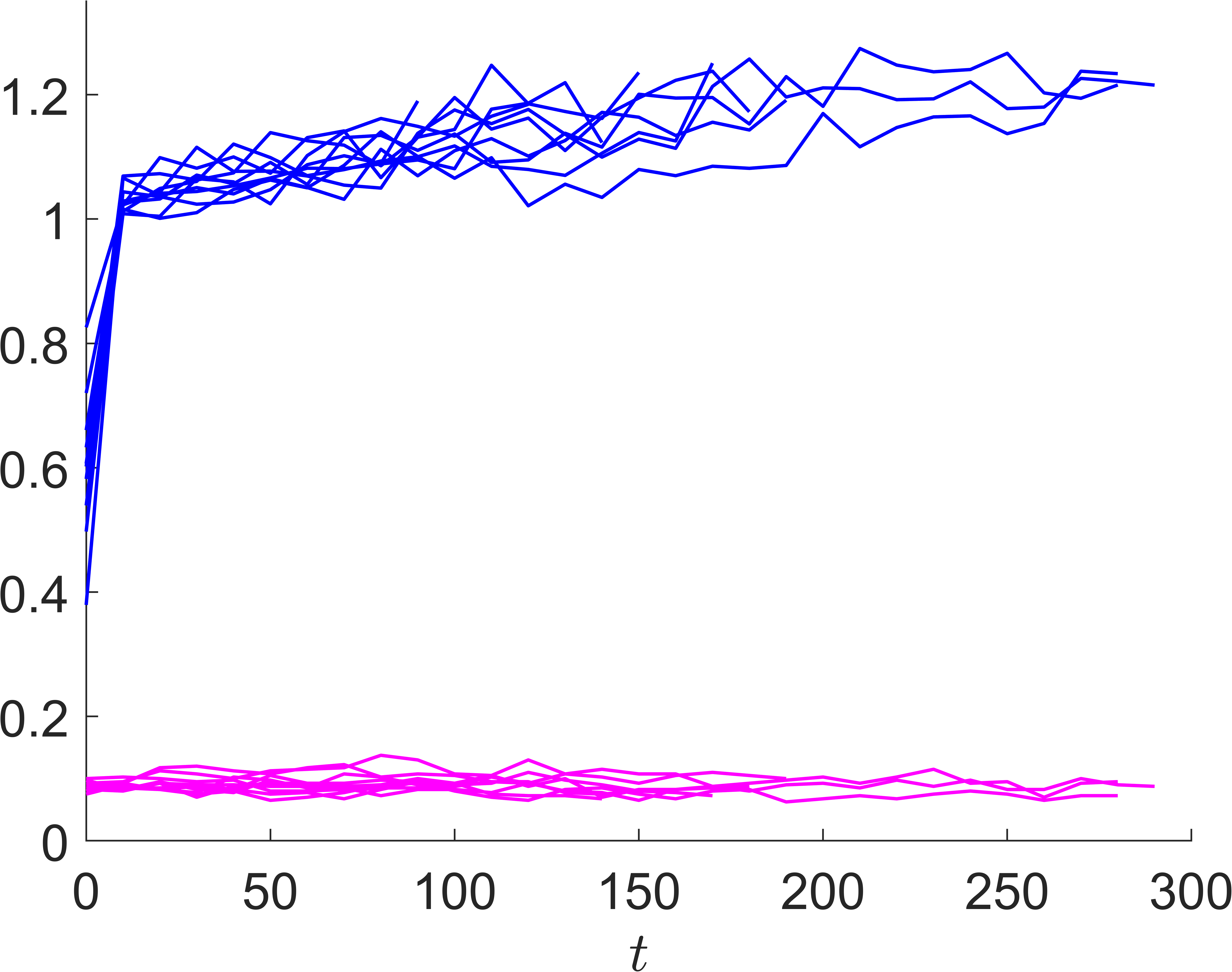}
    \put(2,85){(e)}
    \end{overpic}

    \caption{Top panels: (a) fraction of contained targets at the end of the simulation  $\chi$ and (b) logarithm of the ratio of the end time of the simulation over the maximum allowed time $\log (t^{end}/t^{max})$. (c-e): $10$ time dynamics of the normalized average herders' distance from the origin  $\langle|\textbf{H}|\rangle(t)/R$ and fraction of contained targets $\chi(t)$ (c) Both $\langle|\textbf{H}|\rangle/R$ and $\chi$ converge to a stationary value. (d) The herders rapidly shepherd all the targets in the goal region $\og$ and the simulation is stopped for $t^{end}<t^{max}$ when all the herders are inside $\og$. (e) The herders are outnumbered and are not able to balance the diffusion of the targets; the simulation is stopped for $t^{end}<t^{max}$. In all the above simulations $R=50$, $\xi=10$, $t^{max}=5.0\times 10^3$.}
    \label{fig:ChiRhoVSTime}
\end{figure*} 

\bibliography{biblioSM}
\newpage